\documentclass[aps,prc,onecolumn,showpacs,amsmath,amssymb,nofootinbib,11pt]{revtex4-2}
\usepackage{graphicx}
\usepackage{color}
\usepackage{times}
\usepackage{inputenc}
\usepackage{bm}
\usepackage{ulem}
\usepackage{multirow}
\usepackage{url}
\usepackage{natbib}
\usepackage{mathrsfs}
\usepackage{physics}
\usepackage{comment}
\usepackage{enumitem}
\usepackage[table,xcdraw]{xcolor}
\usepackage{booktabs,makecell}
\usepackage{comment}
\usepackage[colorlinks=true,citecolor=blue,urlcolor=blue,linkcolor=blue]{hyperref}
\renewcommand{\l}{\left}
\renewcommand{\r}{\right}
\usepackage{orcidlink}

\begin{document}

\title{General relativistic study of $f$-mode oscillations in neutron stars with gravitationally bound dark matter}

\author{Pinku Routaray\,\orcidlink{0000-0002-6746-7719}}
\email{routaraypinku@gmail.com}
\affiliation{\it Department of Physics and Astronomy, National Institute of Technology, Rourkela 769008, India}

\date{\today}

\begin{abstract}
A comprehensive investigation of nonradial oscillations in neutron star (NS) admixed with gravitationally bounded dark matter (DM) is carried out within the framework of full general relativity. The relativistic mean field (RMF) formalism is employed to illustrate the hadronic equation of state (EOS), while a physically motivated, gravitationally captured, non-uniform fermionic Higgs-portal DM component is incorporated to model DM-admixed NS. The DM distribution is characterized by two free parameters: $\alpha M_\chi$, an effective control parameter that combines the DM concentration and the DM candidate mass, and $\beta$, a steepness parameter controlling the DM density distribution. The quasi normal mode (QNM) characteristics such as fundamental ($f$) mode frequency and its corresponding gravitational-wave (GW) damping time ($\tau$) is calculated for DM-admixed NS by solving the general relativistic perturbed equations involving axial as well as polar modes. The study demonstrates how the inclusion of DM distribution modifies the $f$-mode frequency and enhances the damping rate, reflecting a stronger coupling between matter and spacetime perturbations. Considering DM effects, the correlation analysis among DM model parameters, NS observables and QNM characteristics also carried out. Analytic fits for the $f-C-\tau$ and $f-\Lambda -\tau$ relations are constructed and calibrated for DM-admixed NS models. Building upon asteroseismic universal relations (URs), multimessenger constraint from the GW170817 event is employed by mapping the tidal deformability $\Lambda_{1.4}$ into the $(f_{1.4},\tau_{1.4})$ space, thereby providing observational bounds on the oscillation properties of canonical DM-admixed NS model.
\end{abstract}
\maketitle

\section{Introduction}

Neutron stars represent one of the most compact states of matter in the universe, characterized by exceptionally high densities. They are formed in the aftermath of core-collapse supernova explosions, when massive stars with initial masses in the range $8-20 \ M_\odot$, where $M_\odot$ denotes the solar mass, undergo gravitational collapse.  Due to their supra-nuclear core densities, typically in the order of $2-10 n_0$, where $n_0$ is the nuclear saturation density, the internal properties of NSs are unique, making them natural laboratories for studying dense matter physics, and these conditions are beyond the reach of current terrestrial experiments \cite{Lattimer_2001, Lattimer_Rev_2012, Lattimer_Rev_2021}. Thanks to recent advancements in multi-messenger astronomy, including X-ray \cite{Miller_2019, Riley_2019, Miller_2021, Riley_2021} and GW observations \cite{Abbott_2017, Abbott_2018}, we now have precise measurements of pulsar masses, radii, and tidal deformabilities during binary NS mergers. These measurements are essential for constraining the NS EOS, which in turn helps us understand the internal composition of these compact objects. NSs primarily consist of baryons, and in addition, their interior may host exotic components such as kaon condensates \cite{Glendenning_kaon_1998, Brown_kaon_2006, Thapa_kaon_2020, FuMa_kaon_2022}, hyperons \cite{Schaffner_hyperon_1996, Weissenborn_hyperon_2012, Bednarek_hyperon_2012, Fortin_hyperon_2015}, deconfined quark matter \cite{Weber_quark_1999, Orsaria_quark_2014, Baym_quark_2018, Annala_quark_2020, Ferreira_quark_2020}, or a quarkyonic phase \cite{McLerran_quarkyonic_2019, Zhao_quarkyonic_2020, Cao_quarkyonic_2022}. Given their extreme environment, another intriguing possibility is the presence of DM, an exotic component that could be accumulated and interact with the NS's interior to further affect their observables. 

DM is believed to be the predominant form of matter that is still concealed in the universe \cite{Bertone_2005}. DM is studied using a variety of methods, including gravitational lensing \cite{Massey_2010}, the cosmic microwave background \cite{Wayne_2002}, and studies of spiral galaxy rotation curves \cite{Begeman_1991}. Despite this multiple and consistent body of evidence, the true nature and particle identity of DM are still mysterious. Understanding DM properties will make it more feasible for observational astrophysics to determine its nature. These observations are made using techniques including particle colliders, astrophysical probes, direct detection, and indirect detection. A unique indirect way to examine DM characteristics is through the NS due to its compact and robust structure \cite{Lavallaz_DM_in_NS_2010, Kouvaris_DM_in_NS_2010, Ciarcelluti_DM_in_NS_2011, Goldman_DM_in_NS_2013, Güver_DM_in_NS_2014, Ellis_DM_in_NS_2018, Bramante_DM_in_NS_2018}. NS's observables, which include their mass, radius, and tidal deformability, undergo substantial changes whenever DM is introduced into their structure \cite{Narain_2006, Panotopoulos_2017, Ellis_2018, Nelson_2019, harishprd_2021, arpan_two_fluid_2022, Davood_2022, Giangrandi_2023, pinku_prd_2023, pinku_jcap_2023, pinku_mnras_2023, Shirke_2024, pinku_ijmpe_2024, pinku_prd_2025, pinku_pdu_2025, Hajkarim_2025}. 

There are two primary approaches to modeling the interaction between DM and NS: (1) gravitational interaction (two-fluid mechanism), (2) non-gravitational interaction (single-fluid mechanism). In the case of gravitational interaction, DM is assumed to interact with a NS solely through gravity, without the exchange of any mediator particles. Such an interaction can lead to different structural configurations, either with DM forming a central core or an extended halo, depending on the DM mass, its fractional abundance inside the NS, and the strength of the gravitational coupling. Several works has been carried out to explain this structure in NS and studied how NS's properties influenced during this phenomena \cite{Nelson_2019, Ivanytskyi_2020, arpan_two_fluid_2022, Davood_2022, Leung22, Miao_2022, Mariani_2023, Rutherford_2023, Diedrichs_2023, Giangrandi_2023, Shakeri_2024, pinku_prd_2025, Marzola_Odilon_2025}. For instance, Nelson {\it et al.} \cite{Nelson_2019} studied the impact of a trace amount of asymmetric DM on gravitational-wave emission and demonstrated how a DM halo can emerge during the inspiral phase. Their analysis compared both fermionic and bosonic DM scenarios. Subsequently, using a two-fluid formalism, Ivanytskyi {\it et al.} \cite{Ivanytskyi_2020} determined the allowed mass range of fermionic asymmetric DM and the corresponding DM mass fraction in the massive pulsars PSR J0348+0432 \cite{Antoniadis13} and PSR J0740+6620 \cite{Fonseca21}. They further showed that light DM particles ($< 0.2 \ {\rm GeV}$) can form an extended halo around the NS, leading not to a decrease but rather to an increase in the total gravitational mass of the star. Bosonic DM was considered in Ref. \cite{Davood_2022}, where the authors demonstrated that the formation of either a core or a halo configuration primarily depends on the mass and fractional abundance of sub-GeV bosons in the strong-coupling regime. Self-interacting fermionic DM was investigated by Miao {\it et al.} \cite{Miao_2022}, who established conditions for dark core or halo formation using Bayesian inference and also examined the effects of dark halos on pulsar pulse profiles. A complementary statistical analysis based on NICER observations was presented by Rutherford {\it et al.} \cite{Rutherford_2023}. In particular, their study focused on scenarios in which all DM is confined within the baryonic radius of the NS, resulting in a compact DM core. More recently, in \cite{Sotani_Ankit_2025}, authors derived new non-radial oscillation modes within the Cowling approximation using a two-fluid formalism. They demonstrated that, in the core region, the DM associated $f$-mode frequencies are higher than those corresponding to normal baryonic matter.

In contrast to the gravitational interaction, the non-gravitational interaction assumes that the DM is fully captured within the NS over astrophysical timescales, but they interact with the nucleons via exchanging mediating particles and directly influence the EOS of the NS. This method is regarded as a single-fluid approach used for the solving DM-admixed NS. Several studies have investigated DM-admixed NS within the framework of the single-fluid approximation \cite{Panotopoulos_2017, Arpan_2019, Quddus_2020, harishprd_2021, Debashree_2021, pinku_prd_2023, pinku_jcap_2023, pinku_mnras_2023, Shirke_2024, pinku_prd_2025}. For instance, in Ref. \cite{Panotopoulos_2017}, author explains how DM affects the realistic equation of state in NSs and he kept the DM fermi momentum constant. This approach subsequently utilized in many studies by keeping DM fermi momentum constant which alternatively choose a constant DM density inside the NS \cite{Arpan_2019, Quddus_2020, harishprd_2021, Flores_2024}. However, the assumption of constant Fermi momentum within a NS is physically inconsistent. Given that NSs are gravitationally bound systems with extremely strong gravity, the DM density should naturally be higher at the core than at the crust or surface. Hence, an important open question concerns the density distribution of DM inside the star, which remains largely uncertain. Since the internal distribution of DM can significantly influence the macroscopic properties and observable signatures of NSs, gaining a deeper understanding of this density profile is essential for constructing realistic models and interpreting astrophysical observations. Recently, Kumar and Sotani \cite{Ankit_2025} addressed the issue of DM distribution in NSs by assuming the DM density as a function of baryon density which reflects a more accurate distribution of DM inside the NS. Motivated with this physical realistic consideration, the present study employs the aforementioned model to derive the impacts of the DM on NS structure and dynamics.

Parallel to progress in static observables, asteroseismology-the study of NS oscillation modes and their GW signatures has matured into a powerful probe of ultra-dense matter. Several quasi normal modes (QNMs) might be present in an oscillating NS, and they are classified based on the force which restores equilibrium \cite{Cowling_1941, Thorne_1967, Kokkotas_1999}. The focus of the current work is on the investigation of the non-radial $f$-mode. Numerous studies in the literature have explored non-radial $f$-mode oscillations of NSs with different degrees of freedom \cite{Yoshida_1997, Sotani_2011, Flores_2014, Sandoval_2018, Wen_2019, Pradhan_2021, harishprd_2021, Zhao_2022, Pradhan_2022, athul_2022, sailesh_2024, Probit_2024, Probit_prc_2024, Sayantan_2024, Dey_2025, Rather_2025}. For instance, Ref. \cite{Pradhan_2021} explored the oscillations of hyperonic NS under the Cowling approximation and later extended the analysis using a full general relativistic treatment \cite{Pradhan_2022}. Their results showed that neglecting metric perturbations leads to an overestimation of the $f$-mode frequency by about $10-30 \ \%$. In another work \cite{Sotani_2021}, various URs were derived within the general relativistic framework, connecting different oscillation modes and damping rates to the dimensionless tidal deformability, providing important tools to connect theoretical predictions utilizing observational astrophysics.

In this work, the non-radial oscillation for DM-admixed NS is investigated. Several previous studies also examined non-radial oscillation in DM-admixed NS \cite{harishprd_2021, Debashree_2021, Flores_2024, Shirke_2024, Thakur_2025, Husain_2026}. For instance, in \cite{harishprd_2021}, the Higgs portal DM model has been employed to compute the non-radial $f$-mode frequency of DM-admixed NS. They have considered the constant DM density in the star without capturing gravitational effects. In Ref. \cite{Flores_2024}, the authors considered the similar Higgs portal DM model with constant DM density to investigate the non-radial oscillation. They have computed the both $f$-mode and damping time using full general relativistic framework. These studies shows that that the presence of DM generally leads to a reduction in mass, radius and tidal deformability and to systematic shifts in the $f$-mode frequency, with magnitude of the effect of total DM fraction rather than its spatial distribution. However, a constant DM density approach is physically unreliable because it does not capture the effect of strong gravity on DM density profile inside the star. Additionally, in \cite{Shirke_2024, Thakur_2025, Husain_2026}, they have considered the self-interacting DM model motivated by neutron decay anomaly and computed the QNMs. They have demonstrated how the DM interaction strength affects the NS observables. Here, the current study refine the previous results by employing a physically motivated Higgs-portal fermionic DM to model the DM-admixed NS which self-consistently incorporates gravitational confinement. The quadrupolar non-radial $f$-mode oscillations and their associated damping times of DM-admixed NSs are studied within a full general relativistic framework, treating both fluid and metric perturbation consistently. Furthermore, the UR for the $f$-mode frequency and $\tau$ as functions of compactness and dimensionless tidal deformability are investigated for DM-admixed NS. Constraints on these quantities for a canonical NS model are further imposed using multimessenger observations, in particular GW data from GW170817 event.

The structure of the paper is organized as follows. Sec. \ref{EOS_DMANS} introduces the formalism used to construct the DM-admixed NS EOS. In Sec. \ref{tov}, the hydrostatic equilibrium structure is discussed, while Sec. \ref{qnm} presents the general relativistic framework for oscillation modes. The results and discussion are provided in Sec. \ref{RD}, followed by the summary and conclusions in Sec. \ref{summary}.

\section{Formalism}

\subsection{Equation of State of Dark Matter Admixed Neutron Star}
\label{EOS_DMANS}
The detailed understanding of NS's EOS decodes its composition. Along with baryon contributions, an exotic component such as DM is considered captured inside the NS during its evolutionary process. There are notable impacts on NS characteristics when a DM core having a mass fraction of around $5\%$ is present within the NS \cite{Ellis_DM_in_NS_2018}. There is no uniform quantity of DM found within the NS; rather, it varies depending on the NS's formation environment and evolutionary history (age, initial temperature). The current study considers the interaction of DM with baryons through the Higgs portal, making DM-admixed NS as a single-fluid approximation. The non-annihilating  Weakly Interacting Massive Particles (Neutralino) is chosen as the DM candidate. 

Here, two separate fluids; baryonic matter and DM are modeled within a single-fluid framework, motivated by the assumption that DM is sufficiently coupled to baryonic matter. Although the DM is treated as collisionless with weak interactions and often approximated as a pressureless component, and also interaction between DM and baryons can lead to efficient thermalization. Through this process, DM can exchange energy and momentum with baryonic fluid, allowing both components to evolve coherently. Under such conditions, the coupling between DM and baryonic matter becomes strong enough to justify describing the system using a single-fluid approximation.


To study the DM-admixed NS, the interaction of baryon and DM is taken within the RMF formalism \cite{FURNSTAHL_1996, Singh_2014, Kumar_2018, pinku_jcap_2023, Probit_2024}. The Hornick3 model is chosen from a series of Hornick models to derive NS properties \cite{Hornick_2018}. To perform the unified EOS treatment, the BPS crust has been used \cite{BPS_crust_1971}. The Hornick3 exhibits a stiff behavior with the maximum-mass $2.52 \ M_\odot$ and canonical radius $R_{1.4}=13.25 \ {\rm km}$.

The DM interaction with baryon can be represented by the following Lagrangian density \cite{Panotopoulos_2017, Quddus_2020, harishprd_2021, pinku_prd_2023},
\begin{eqnarray}
{\cal{L}}_{\rm DM} & = & \bar \chi \left[ i \gamma^\mu \partial_\mu - M_\chi + y h \right] \chi +  \frac{1}{2}\partial_\mu h \partial^\mu h  \nonumber\\
& &
- \frac{1}{2} M_h^2 h^2 + f \frac{M_n}{\nu} \bar \varphi h \varphi \, , 
\label{eq:LDM}
\end{eqnarray}
Here, the $\chi$ and $h$ represent the DM fermion and Higgs field associated with their mass $M_\chi$ and $M_h$, respectively. The $\psi$ represents the standard model nucleon field and has its mass $M_n$. The quantity $\nu$ denotes the Higgs vacuum expectation and it is set as $246 \ {\rm GeV}$. The numerical value of $y$ is kept as $0.06$, which is the coupling constant between DM and the standard model Higgs field. The effective Yukawa coupling involving Higgs and nucleon field is denoted by $fM_n/\nu$, and its value is maintained at $1.145\times10^{-3}$.

Numerous studies have been explored earlier for the DM-admixed NS with Higgs portal DM \cite{Panotopoulos_2017, Quddus_2020, harishprd_2021, pinku_jcap_2023, Flores_2024}. These studies assume that the uniform DM density throughout the NS is considered constant DM Fermi momentum ($k_f^{\rm DM}$). Extending this model to a more realistic approach in Ref. \cite{Ankit_2025}, the author considers the density distribution in the NS uniform, rather than it peaks at the core and diminishes towards the outer layers. The DM density can be expressed as follows,
\begin{eqnarray}
    \frac{n_{\rm DM}}{n_0}=\alpha\Big(\frac{n_{\rm B}-n_t}{n_0}\Big)^\beta
    \label{eq:den_DM}
\end{eqnarray}
Here $n_{\rm DM}$ and $n_{\rm B}$ represent the DM and baryon number density, with $n_0$ as nuclear saturation density without DM. The quantity $n_t$ represents the core-crust transition density within NS. Two independent dimensionless parameters$-$ $\alpha$ and $\beta$ have been introduced in the equation used to regulate the behavior as well as physical distribution of DM.

\begin{enumerate}[label=(\roman*)]
    \item $\alpha$ denotes the scaling factor  that scales the DM density with respect to nuclear saturation density. In practice, it specifies how much DM is present compared to baryonic matter at a given baryon density.
    \item  $\beta$ defines the steepness parameter. It controls how quickly the DM density rises with baryon density. Larger values of $\beta$ make the DM density grow much more sharply, especially as we move toward the core of the star. 
\end{enumerate}

These two arbitrary constants are introduced to maintain the variable DM density which is consistent with the gravitational trapping of DM inside the NS. Gravitational capture naturally leads to a higher DM concentration near the core compared to the crust. Since the current study assumes that DM remains confined within the stellar core, the DM density is set to zero for baryon densities below the core-crust transition density, $n_B \leq n_t$ reflecting the idea that the star's gravitational field traps DM in the core region, preventing it from occupying the outer layers. For the crust region, we utilize BPS crust \cite{BPS_crust_1971}.

The parameters $\alpha$ and $\beta$ enter the equilibrium equations through Eq. \ref{eq:den_DM}, which is used to derive the DM EOS. In deriving the EOS for DM admixed NS, we also enforce beta equilibrium and charge neutrality so that the matter inside the NS remains in full chemical equilibrium, as required for a stable configuration. The variation of $\alpha$ and $\beta$ directly affects the total energy density of the DM-admixed NS. In Ref.\cite{Ankit_2025}, it was shown that, at constant $\beta$, the total energy density is effectively governed by $\alpha M_\chi$.
\begin{itemize}
    \item $\alpha M_\chi$ is the effective control parameter, where $M_\chi$ is the mass of the DM particle.
\end{itemize}
This combination yields identical results, regardless of the individual values of $\alpha$ and $M_\chi$, highlighting $\alpha M_\chi$ as the relevant effective control parameter. Hence in the current study, $\beta$ and $\alpha M_\chi$ taken as the free parameters to govern the DM effects on NS properties.

Now, we can write the DM energy density and pressure as follows \cite{Arpan_2019},
\begin{eqnarray}
{\cal{E}}_{\rm DM} = \frac{2}{(2\pi)^{3}}\int_0^{k_f^{\rm DM}} d^{3}k \sqrt{k^2 + (M_\chi^\star)^2 } + \frac{1}{2}M_h^2 h_0^2 \, ,
\label{eq:edm}
\end{eqnarray}
\begin{eqnarray}
P_{\rm DM} = \frac{2}{3(2\pi)^{3}}\int_0^{k_f^{\rm DM}} \frac{d^{3}k \hspace{1mm}k^2} {\sqrt{k^2 + (M_\chi^\star)^2}} - \frac{1}{2}M_h^2 h_0^2 \, ,
\label{eq:pres}
\end{eqnarray} 
Here, $M^\star$ denotes the DM effective mass, $M^\star=M_\chi-yh_0$. 

Consequently, the total energy density and pressure for the DM-admixed NS may be expressed as \cite{Arpan_2019}, 
\begin{eqnarray}
{\cal{E}}={\cal{E}}_{\rm NS}+ {\cal{E}}_{\rm DM} \, ,
\nonumber
\\
{\rm and}
\hspace{1cm}
P=P_{\rm NS} + P_{\rm DM} \, .
\label{eq:EOS_total}
\end{eqnarray}
where ${\cal{E}}_{\rm NS}$ and $P_{\rm NS}$ represents the energy density and pressure without DM respectively.

After having the EOS of DM admixed NS in hand, we can use the hydrostatic equilibrium equations to obtain the mass-radius profile. The detail of hydrostatic equilibrium equations with boundary conditions is presented in Sec. \ref{tov}. Also, with the DM-admixed NS EOS, we have studied the non-radial oscillation of the star and analyze the $f$-mode frequency and damping time, presented in Sec. \ref{qnm}. Before studying the current Higgs portal DM model with NS in a single fluid mechanism, we need to justify the physical modeling of the DM-admixed NS as a single fluid system.

The physical modeling of single fluid approximation can be justified by scanning the parameter space for the experimental evidence by keeping limits of scattering cross section \cite{Bhat_2020}. \\
A key component of this single fluid model is the DM-Higgs coupling factor ($y$). Its strength can be limited by various DM detection experiments. Despite the fact that the direct detection experiment has not yet shown any events, they did offer some upper bounds on the WIMP-nucleon scattering cross section. Thus, in the current model, the Higgs is mediated while DM is interacted with the nucleons, as a result , there is an elastic scattering between WIMPs and nuclei primarily at the quark level. Given this, the following is an expression for interaction Lagrangian that incorporates the DM wave function ($\chi$) and quark wave function ($q$) with the scalar operator $\bar{\chi}\chi\bar{q}q$, 
\begin{eqnarray}
    \mathcal{L}_{\rm int} = \alpha_q \bar{\chi}\chi\bar{q}q,
\end{eqnarray}
where $\alpha_q = \frac{yfm_q}{\nu M_h^2}$. \\
Now considering the contribution of the above scalar operator, the spin independent scattering cross-section may be expressed as, 
\begin{eqnarray}
    \sigma_{\rm SI} = \frac{y^2f^2M_n^2}{4\pi} \frac{\mu_r^2}{\nu^2M_h^4},
\end{eqnarray}
Here $\mu_r = \frac{M_nM_\chi}{M_n+M_\chi}$ is the reduced mass.\\
In the current study, we considered $y=0.06$, $M_\chi = 100 \ {\rm GeV}$, $f=0.35$, $M_h=125 \ {\rm GeV}$ and $\nu = 246 \ {\rm GeV}$ and obtain $\sigma_{\rm SI} = 7.05 \times10^{-46} \ {\rm cm^2}$. \\
This value is consistent with direct detection experiments such as XENON-1T \cite{Xenon1T_2016}, PandaX-II \cite{PandaX_2016}, and LUX \cite{LUX_2017} with $90\%$ confidence level. Additionally, the current model also satisfies the limits of LHC which is in the range of $10^{-40}$ to $10^{-50} \ {\rm cm^2}$ \cite{Djouadi_LHC_2012}.
\\
Now, it is crucial to verify whether the DM-baryon interaction strength supports single fluid oscillation or any co-moving behavior is raised which may leads to two fluid treatment. To justify the single fluid oscillation, we estimate relaxation time ($\tau_{\rm relax}$) and it needs to be shorter than the NS's oscillation ($\tau_{\rm osc}$).

Now, $\tau_{\rm relax} \sim \frac{1}{n_B \sigma_{\rm SI} v}$. Where $n_B=\rho / M_n$. The typical core density, $\rho \sim 10^{15} \ g/cm^3$ and $M_n = 1.673\times10^{24} \ g$. The characteristic relative velocity was taken as $v \sim 0.1c$, consistent with the order of magnitude of the neutron Fermi velocity expected in dense NS core. Utilizing these values, we can estimate the relaxation time which is in the order of millisecond period of the NS oscillation. The condition $\tau_{\rm relax} < \tau_{\rm osc}$ alone does not rigorously guarantee a single-fluid description. The present analysis is performed within an effective co-moving or strong coupling regime, where the DM and baryonic components are assumed to oscillate approximately together. The present framework does not include relative fluid motion, independent DM oscillation modes, or dynamical two-fluid couplings, which may become important for weaker interactions or sufficiently large self-gravitating DM cores. 


\subsection{Hydrostatic Equilibrium}
\label{tov}
To explain the hydrostatic equilibrium of NS, the Tolman-Oppenheimer-Volkoff (TOV) equations are inferred from Einstein's field equations in Schwarzschild-like co-ordinates. They are given by \cite{Tolman_1939, Oppenheimer_1939}, 
\begin{align}
\frac{dP}{dr} &=  \frac{-1}{r} \frac{\left( P+\mathcal{E} \right) \left( m + 4\pi r^3P\right)}{\left(r-2m\right)},
\nonumber \\
\frac{dm}{dr} &= 4\pi r^2 \mathcal{E} .
\end{align}

The coupled differential equation can be solved using the initial condition $m(r=0)=0$ and $P(r=0) = P_c$, where $P_c$ denotes the central pressure. The integration proceeds until the surface boundary, at which $m(r=R)=M$ and $P(r=R)=0$.

Alongside the TOV equations, a complementary set of differential equations might be solved to determine the tidal love number $k_2$ for a particular EOS.  This procedure also yields the dimensionless tidal deformability, one more significant observable parameter ($\Lambda$) \cite{Hinderer_2008, Kumar_2017, Flanagan_2018},
\begin{eqnarray}
    \Lambda=\frac{2}{3}k_2\Big(\frac{R}{M}\Big)^5.
    \label{eq:tidal}
\end{eqnarray}

Here, the expression for $k_2$ can be written as, 
\begin{align}
k_2= & \frac{8}{5} C^5(1-2 C)^2\left[2\left(y_2-1\right) C-y_2+2\right] \times\left\{\left[\left(4y_2+4\right) C^4+\left(6 y_2-4\right) C^3-\left(22 y_2-26\right) C^2 \right.\right] \nonumber \\
& \left.+3\left(5 y_2-8\right) C-3\left(y_2-2\right)\right]2 C  +3(1-2 C)^2 \left.\times\left[2\left(y_2-1\right) C-y_2+2\right] \log (1-2C)\right\}^{-1}\,.
\label{eq:k2}
\end{align}

 $y_2$ can be calculated by solving the following differential equation with initial boundary condition $y(0)=2$,
\begin{equation}
    r\frac{dy_2(r)}{dr} + y_2(r)^2 + y_2(r)F(r) + r^2 Q(r) = 0,
\label{eq:tidal_differential}
\end{equation}
Where,
\begin{align}
    F(r) & = \frac{r - 4\pi r^3 \qty{\varepsilon(r) - P(r)}}{r - 2M(r)}, \label{eq:36}\\
    Q(r) & = \frac{4\pi r \qty{5\varepsilon(r) + 9P(r) + \frac{\varepsilon(r) + P(r)}{\pdv*{P(r)}{\varepsilon(r)}}}}{r-2M(r)} - 4 \bqty{\frac{M(r) + 4\pi r^3 P(r)}{r^2\qty{1-2M(r)/r}}}. \label{eq:37}
\end{align}

\subsection{Quasinormal Modes}
\label{qnm}
GW frequencies are obtained by solving an eigenvalue problem based on the static, spherically symmetric NS models. The resulting frequencies are generally complex, with the real part representing the star's oscillation frequency ($f$-mode) and the imaginary part indicating the rate at which these oscillations decay ($\tau$). Such complex frequencies are commonly referred to as QNM,
\begin{eqnarray}
    \omega=2\pi f+i\frac{1}{\tau}.
\end{eqnarray}
Numerous studies have already studied the $f$-mode using the relativistic Cowling approximation while ignoring metric perturbation \cite{Pradhan_2021, harishprd_2021, sailesh_2024, pinku_jcap_2023, Probit_2024}.  However, other research \cite{Sotani_2021, Zhao_2022, Pradhan_2022, Shirke_2024} uses a general relativistic approach to study the mode frequency.  Based on the aforementioned investigations, it can be concluded that the Cowling approximation overvalues the $f$-mode frequency by around $10-30\%$ relative to the frequency determined using a general relativistic approach.

In the current study, the general relativistic framework was employed to calculate the QNM frequency. Considering the appropriate boundary conditions similar to Detweiler and Lindblom \cite{Lindblom_1983}, the perturbation equations have been solved to compute the QNM frequency. At the center of the star, the regularity condition has been enforced for all perturbation variables. Near the center, $r=0$, the expansion of all necessary variables has been done using Taylor power series. The second boundary condition is at the stellar surface, pressure perturbation vanishes. Finally, we have imposed the last boundary condition for the outside perturbation of the star, there exists only outgoing gravitational waves at infinity and Zerilli equation is used to solve this \cite{Zerilli_1970}.  In the current section, we have shown the basic equations with boundary conditions which need to be solved to obtain the complex QNM frequencies.
 
\subsubsection{Perturbation within the star}
The fundamental equations required to determine the complex QNM frequencies are outlined below.
We can write the perturbed metric ($ds^2_p$) as follows \cite{Thorne_1967},
\begin{eqnarray}
    ds^2_p=ds^2+h_{\mu \nu} dx^{\mu}dx^{\nu}~.
    \label{eqn:perturbedmetric}
\end{eqnarray}
Based on the formulation of Thorne and Campolattaro \cite{Thorne_1967}, we consider the even-parity (polar) perturbations, wherein the GW and matter perturbations are coupled. Accordingly, $h_{\mu \nu}$ is expressed as \cite{Thorne_1967, Sotani_2001},
\begin{eqnarray}
    h_{\mu \nu}=
   \begin{pmatrix}
r^lHe^{2\Phi} & i\omega r^{l+1} H_1 &0&0\\
i\omega r^{l+1} H_1 & r^l H e^{2\lambda} &0 &0\\
0 &0 & r^{l+2}K& 0\\
0 & 0& 0 &  r^{l+2}K sin^2{\theta}
\end{pmatrix} Y^l_m e^{i\omega t} ~, \nonumber\\
 \text{ }
 \label{eqn:metricfunctions}
\end{eqnarray}
Here, $Y_l^m$ denote the spherical harmonics, and $H$, $H_1$, and $K$ are the perturbed metric functions, each depending on the radial coordinate $r$ (i.e., $H=H(r)$, $H_1=H_1(r)$, and $K=K(r)$). Using $\boldsymbol{\zeta}=(\zeta^r,\zeta^{\theta},\zeta^{\phi})$, the Lagrangian displacement vector connected to the fluid's polar disturbances may be described as \cite{Lindblom_1983, Tonetto_2021},
\begin{eqnarray}
    \zeta^{r}&=&\frac{r^l}{r}e^{- \lambda} W(r)  Y^l_m e^{i\omega t} \nonumber \\
    \zeta^{\theta}&=&\frac{-r^l}{r^2} V(r)  \frac{\partial Y^l_m}{\partial \theta} e^{i\omega t} \nonumber \\
    \zeta^{\phi}&=&\frac{-r^l}{r^2 sin^2\theta} V(r)  \frac{\partial Y^l_m}{\partial \phi} e^{i\omega t}
    \label{eqn:pertfluid}
\end{eqnarray}
Here $W$ and $V$ represents the amplitudes of the radial and transverse fluid perturbations, respectively. The equations governing these perturbations The metric perturbations inside the star and the equations regulating such perturbation functions are presented by \cite{Sotani_2001, Tonetto_2021},
\begin{eqnarray}
    \frac{d H_1}{dr}&=&\frac{-1}{r}\left[l+1+\frac{2m}{r}e^{2\lambda}+4\pi r^2e^{2\lambda} \left( p-\epsilon \right)\right] H_1 \nonumber\\
    &+&\frac{1}{r}e^{2\lambda}\left[H+K+16\pi\left(p+\epsilon\right)V\right] \label{eqn:dh1} \ , \\
    \frac{d K}{dr}&=&\frac{l\l(l+1\r)}{2r}H_1+\frac{1}{r}H-\l(\frac{l+1}{r}-\frac{d\Phi}{dr}\r)K \nonumber \\
    &+&\frac{8\pi}{r}\l(p+\epsilon\r)e^{\lambda} W \ ,  \label{eqn:dk} \\
    \frac{d W}{dr}&=&re^{\lambda}\l[\frac{1}{\gamma p}e^{-\Phi}X-\frac{l\l(l+1\r)}{r^2}V-\frac{1}{2}H-K\r] \nonumber \\
    &-&\frac{l+1}{r}W   \label{eqn:dw} \ , \\
    \frac{d X}{dr}&=& \frac{-l}{r}X+\l(p+\epsilon \r)e^{\Phi}\Bigg[\frac{1}{2}\l(\frac{d\Phi}{dr}-\frac{1}{r}\r)H \nonumber\\
    &-&\frac{1}{2}\l( \omega^2re^{-2\Phi}+\frac{l(l+1)}{2r}\r)H_1+\l(\frac{1}{2r}-\frac{3}{2}\frac{d\Phi}{dr}\r)K \nonumber\\
    &-&\frac{1}{r}\l[ \omega^2\frac{e^{\lambda}}{e^{2\Phi}}+4\pi \l(p+\epsilon \r) e^{\lambda}-r^2\frac{d}{dr}\l( \frac{e^{-\lambda}}{r^2}\frac{d \Phi}{dr}\r)\r]W \nonumber \\
    &-&\frac{l(l+1)}{r^2}\frac{d\Phi}{dr}V\Bigg]  \label{eqn:dx}\ ,
\end{eqnarray}
\begin{eqnarray}
   &&\l[1-\frac{3m}{r}-\frac{l(l+1)}{2}-4\pi r^2p\r]H-8\pi r^2 e^{-\Phi}X \nonumber\\
   &-& \l[ 1+ \omega^2r^2e^{-2\Phi} -\frac{l(l+1)}{2}-(r-3m-4\pi r^3p)\frac{d\Phi}{dr}\r]K \nonumber \\
   &+&r^2e^{-2\lambda}\l[\omega^2e^{-2\Phi}-\frac{l(l+1)}{2r}\frac{d\Phi}{dr}\r]H_1 =0  \label{eqn:h}\\
  && e^{2\Phi}\l[ e^{-\phi}X+\frac{e^{-\lambda}}{r}\frac{dp}{dr} W+\frac{(p+\epsilon)}{2}H\r] \nonumber \\
  &-&\omega^2 \l(p+\epsilon\r) V=0 ~, \label{eqn:v}
\end{eqnarray}
where $X$ is introduced as\cite{Lindblom_1983, Sotani_2001}
\begin{eqnarray}
    X&=&\omega^2\l(p+\epsilon \r) e^{-\Phi} V-\frac{We^{\Phi-\lambda}}{r}\frac{dp}{dr}-\frac{1}{2} \l(p+\epsilon \r) e^{\Phi}H\,, \nonumber \\
    \text{}  \label{eqn:x}
\end{eqnarray}
The enclosed mass of the star is $m=m(r)$, and the adiabatic index is $\gamma$, which is defined as
\begin{equation}
    \gamma=\frac{\l(p+\epsilon \r)}{p}\l(\frac{\partial p}{\partial \epsilon }\r)\bigg|_{ad} ~.
     \label{eqn:gamma}
\end{equation}
To solve the differential equations [Eqs. \eqref{eqn:dh1}–\eqref{eqn:dx}] along with the algebraic relations [Eqs. \eqref{eqn:h}–\eqref{eqn:v}], appropriate boundary conditions are required. The perturbation functions must remain finite throughout the stellar interior, particularly at the center ($r=0$), and the perturbed pressure ($\Delta p$) must vanish at the stellar surface. The central values of the perturbation functions are obtained using a Taylor series expansion following Appendix B of \cite{Lindblom_1983} (see also Appendix A of \cite{Sotani_2001}). It should be noted that the first term on the right-hand side of Eq. (A15) in \cite{Sotani_2001} is missing a factor of $\epsilon$. The surface boundary condition, $\Delta p = 0$, is equivalent to $X(R)=0$, since $\Delta p = -r^l e^{-\Phi} X$. The numerical procedure developed by Lindblom and Detweiler \cite{Lindblom_1983} is employed to determine the unique solution for a given $l$ and complex frequency $\omega$ that satisfies all boundary conditions within the star.
\subsubsection{Perturbations outside the star and complex eigenfrequencies}
The Zerilli equation determines the perturbations outside the star\cite{Zerilli_1970}.
\begin{equation}
    \frac{d^2Z}{dr_*^2}+\omega^2 Z=V_Z Z
    \label{eqn:zerilli}
\end{equation}
 Here $r_*=r+2M \log \l({\frac{r}{2M}-1}\r)$ represents the tortoise co-ordinate and $V_Z$ which can be written as \cite{Zerilli_1970},
 \begin{eqnarray}
     V_Z&=&\frac{2\l(r-2M\r)}{r^4 \l(nr+3M\r)^2}\Big[n^2(n+1)r^3 \nonumber \\
     &+&3n^2Mr^2+9nM^2r+9M^3\Big] ~,
 \end{eqnarray}
where $n=\frac{1}{2} (l+2)(l-1)$. It is possible to describe the wave solution to \eqref{eqn:zerilli} asymptotically as \eqref{eqn:zerillisolution},
\begin{eqnarray}
    Z=A(\omega)Z_{in}&+&B(\omega) Z_{out}\,, \label{eqn:zerillisolution}\\
    Z_{out}=e^{-i\omega r^*} \sum_{j=0}^{j=\infty}\alpha_j r^{-j}&,& Z_{in}=e^{i\omega r^*} \sum_{j=0}^{j=\infty}\bar{\alpha}_j r^{-j} ~.\nonumber
\end{eqnarray}
Keeping  terms up to $j=2$ one finds,
 \begin{eqnarray}
     \alpha_1&=&-\frac{i}{\omega}(n+1)\alpha_0, \\
     \alpha_2&=&\frac{-1}{2\omega^2}\l[n(n+1)-i3M\omega\l(1+\frac{2}{n}\r)\r]\alpha_0
 \end{eqnarray}
The approach outlined in \cite{Lindblom_1983, Sotani_2001, Fackerell_1971} is used for initial boundary values of Zerilli functions.  Once $m=M$ and the perturbed fluid variables outside the star are set to 0 (i.e., $W=V=0$), the relationship between the metric functions \eqref{eqn:metricfunctions} and the Zerilli function ($Z$ in Eq.\eqref{eqn:zerilli}) may be expressed as follows,
\begin{eqnarray}
    \begin{pmatrix}
    r^lK\\
    r^{l+1}H_1
    \end{pmatrix}
    &=&Q\begin{pmatrix}
    Z \\
    \frac{dZ}{dr_*}
    \end{pmatrix}
    \label{eqn:zerillconnection}
\end{eqnarray}
\begin{eqnarray*}
    Q&=&\begin{pmatrix}
    \frac{ n(n+1)r^2+3nMr+6M^2}{r^2(nr+3M)} & 1\\
    \frac{nr^2-3nMr-3M^2}{(r-2M)(nr+3M)} & \frac{r^2}{r-2M}
    \end{pmatrix} \\
\end{eqnarray*}
Zerilli functions' initial boundary values have been set using \eqref{eqn:zerillconnection}. After that, the Zerilli equation \eqref{eqn:zerilli} is numerically integrated to infinity, yielding complex coefficients $A(\omega)$ and $B(\omega)$ that match the analytic expressions for $Z$ and $\frac{dZ}{dr_*}$ together with the numerically calculated values of $Z$ and $\frac{dZ}{dr_*}$. The natural oscillation frequencies of a NS, arising in the absence of any external driving by incident GWs, are referred to as its QNM frequencies. In order to describe the complex eigenfrequencies of QNMs mathematically, we must obtain the complex roots of $A(\omega)=0$. In the exterior region, the solution is integrated to a sufficiently large radial distance ($r \sim 50 \omega^{-1}$), where the asymptotic form of the wave is well defined.

All relevant ODEs were numerically integrated using an adaptive explicit Runge-Kutta method of order 5(4) (RK45). Throughout the integration, we have maintained both absolute and relative tolerances in the range of $10^{-10}$, ensuring stable and convergent solutions. Numerical convergence was verified by varying the tolerance and confirming that the resulting physical outcomes remain unchanged within the given precision. The numerical error used while searching $\omega = \omega_R + i\omega_I$ was $|\Delta \omega| < 10^{-5}$. By comparing our results with those in Refs. \cite{Sotani_2021, Pradhan_2022, athul_2022}, the correctness of our code was carefully examined.

\section{Results And Discussions}
\label{RD}
 
As discussed earlier, the DM interaction with nucleonic matter is considered as a single-fluid, where the EOS is directly impacted by the DM. Here DM is gravitationally captured and distributed non-uniformly inside the star. This phenomenon, in turn, affects properties of the NS such as mass, radius, compactness, tidal deformability, and complex QNM frequencies. Hence, all the numerical outcomes pertaining to the DM-admixed NS properties are presented in this section. Additionally, the URs in NS asteroseismology are also investigated by considering DM within the NS.

\subsection{Distribution of dark matter}
The current model rely on the assumption that the captured DM distributed non-uniformly inside the NS. This distribution could also affect the baryonic matter inside the NS. Hence, here we have studied the radial profile of the DM and baryonic matter.
\begin{figure}
    \centering
    \includegraphics[width=1.0\linewidth]{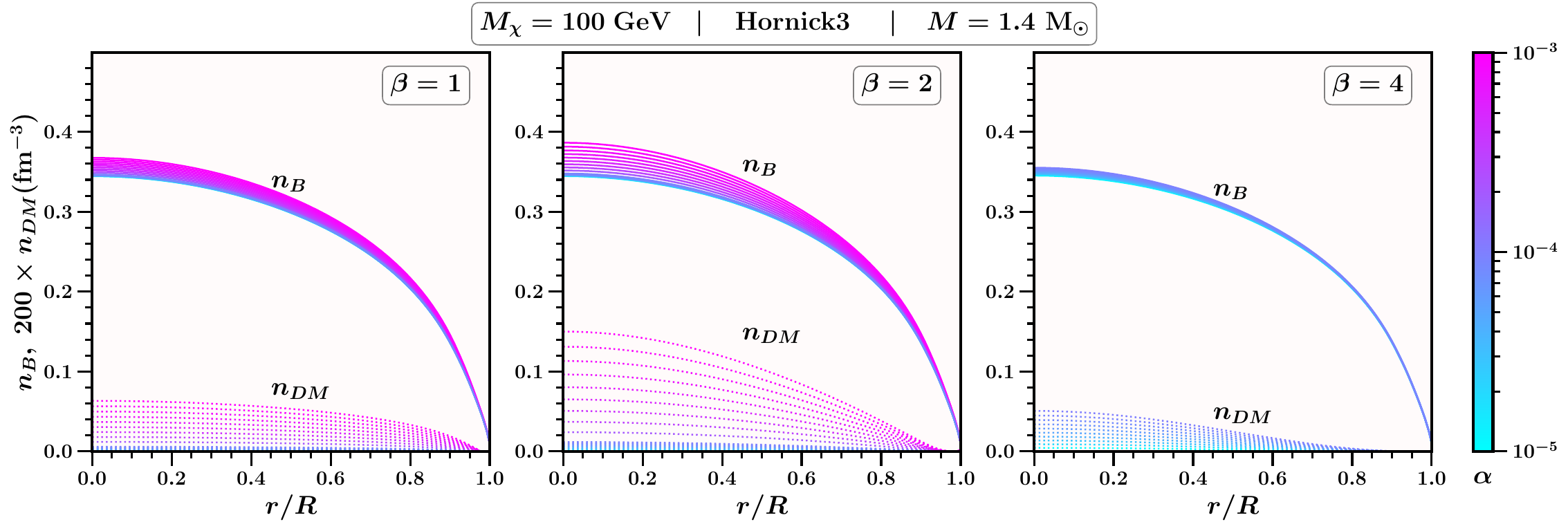}
    \caption{The radial profile of DM density ($n_{DM}$), and baryon density ($n_B$) for $1.4 \ M_\odot$ NS with Hornick4 model. For the $\beta = 1, {\rm and} \ 2$, the $\alpha$ value varies from $10^{-5}$ to $10^{-3}$, where as for $\beta=4$, it varies from $10^{-5}$ to $10^{-4}$.}
    \label{fig:density_profile}
\end{figure}
\begin{figure}
    \centering
    \includegraphics[width=1.0\linewidth]{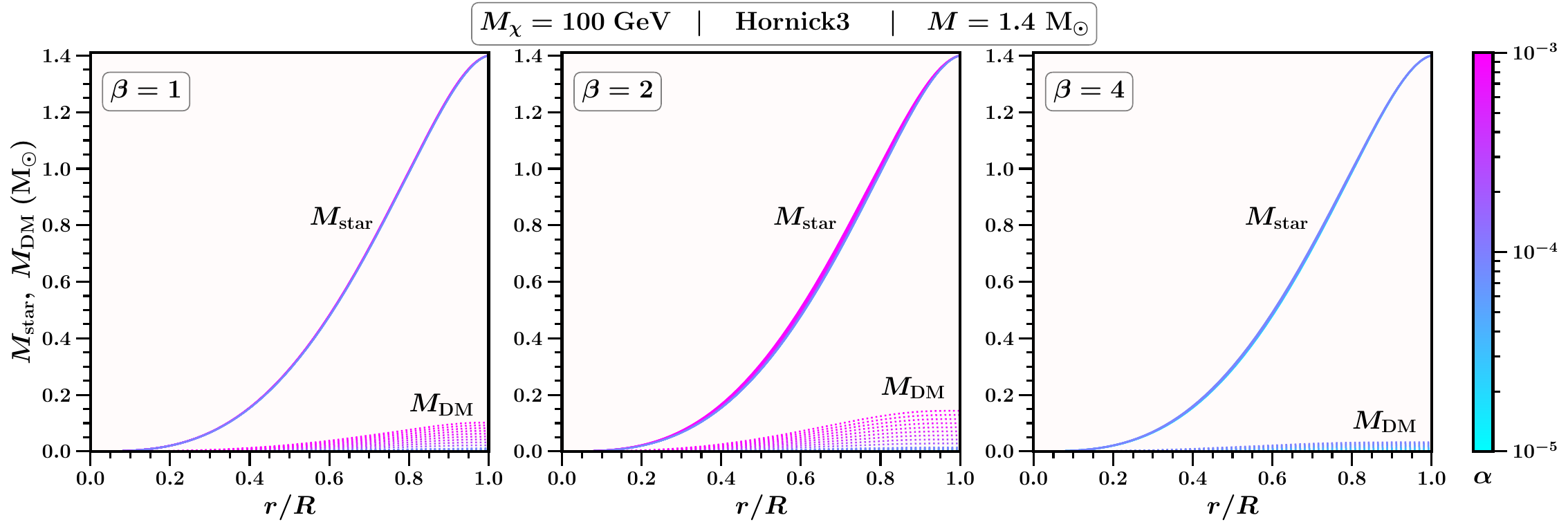}
    \caption{Same as Fig. \ref{fig:density_profile}, for mass.}
    \label{fig:mass_profile}
\end{figure}

In Fig. \ref{fig:density_profile}, the radial density profile for both DM and baryon density is shown. With the several values of $\alpha$, which ranges from $10^{-5}$ to $10^{-3}$. The $\beta$ has taken three distinct value: 1 (left panel), 2 (middle panel), and 4 (right panel). Here, we maintain a constant DM particle mass, $M_\chi=100$ GeV for all the values of $\alpha$ and $\beta$. The density profile for $n_B$ without DM is presented in every panel for comparison with the changes due to DM. To make it easier to figure out, the $n_{DM}$ is multiplied by 200, as shown by the y-axis label.
\\
It is observed that the concentration of DM inside the NS becomes higher when $\alpha$ increases, despite the overall NS's mass kept the same at $1.4 \ M_\odot$.  This behavior is consistent with the corresponding functional form, $\frac{n_{\rm DM}}{n_0}=\alpha\Big(\frac{n_{\rm B}-n_t}{n_0}\Big)^\beta$, whereby $\alpha$ regulates the amplitude of DM density and $\beta$ determines the steepness. When $\alpha$ increase, $n_{DM}$ increases up to very smaller percentage compared to the $n_B$, still influence the NS structure in significant manner. When $\beta=1$, a more even distribution of DM density can be seen across the core, but the distribution suddenly peaks for higher value of $\beta=4$ at the center, which shows that there is strong gravitational confinement. The higher DM concentration makes the NS core's gravitational potential stronger, which means that baryonic matter has to redistribute to maintain hydrostatic equilibrium. As DM's gravitational pull becomes stronger, baryons gravitate toward the core to balance off the stronger core potential. This behavior shows how baryonic matter changes to be stable when DM's gravitational pull is greater, even if the overall mass stays the same. A similar behavior also can be observed in the Fig. \ref{fig:mass_profile}, where we show the radial distribution of mass of the DM ($M_{DM}$), mass of the star ($M_{\rm star}$). The mass of the DM is increases with $\alpha$ as well as $\beta$, while maintaining the total mass of the star at $1.4 \ M_\odot$.
\\
This modification of the core structure directly affects the structural properties as well as the asteroseismic properties. In particular, the increased central concentration enhances the effective compactness and GW emission efficiency, which explains the observed increase in $f$-mode frequency and reduction in the damping time, $\tau$. Therefore, the changes in $f$ and $\tau$ can be physically traced back to the profile-dependent redistribution of mass and density inside the star.

\subsection{Mass-radius relations}
\begin{figure*}
   \centering
   \includegraphics[width = 1.0\textwidth]{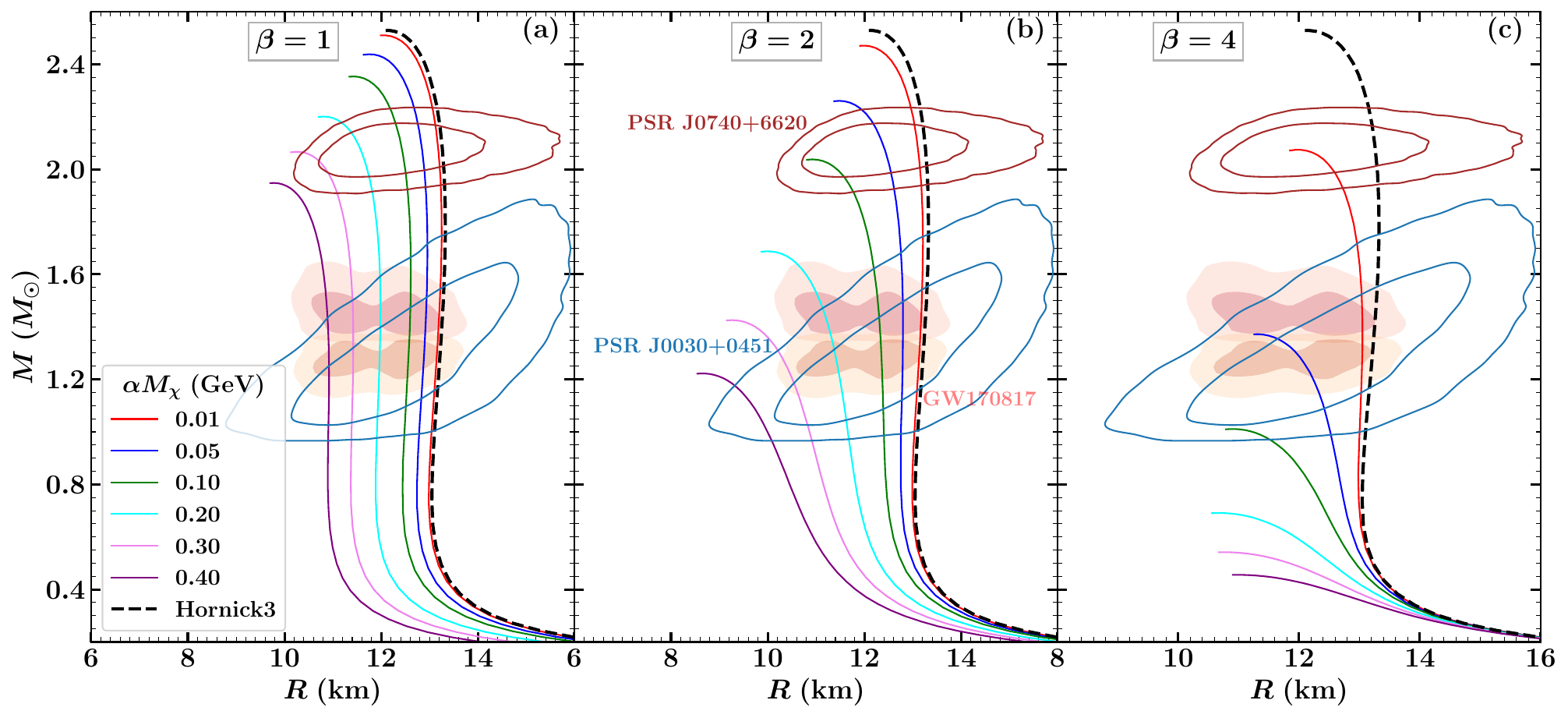}
   \caption{The mass-radius relation of DM-admixed NS is shown for Hornick3 EOS \cite{Hornick_2018}. Different choices of $\alpha M_\chi$ parameter are compared at different steepness parameter in different panels; $\beta=1$ (a), $\beta=2$ (b) and $\beta=4$ (c). The $1\sigma$ and $2\sigma$ mass-radius constraints from PSR J0030+0451 \cite{Riley_2019} and PSR J0740+6620 \cite{Riley_2021} are used. Also, the $50\%$ and $90\%$ confidence intervals from the LIGO-Virgo analysis for binary NS components of the GW170817 event are imposed \cite{Abbott_2017}.}
    \label{fig:mr}
\end{figure*}
In Fig. \ref{fig:mr}, at different values of $\beta$, the mass-radius relations are plotted with varying $\alpha M_\chi$ for Hornick3 EOS. For the Hornick3 EOS with a BPS crust, the transition density is given by, $n_t=5.1\times10^{-2} \ {\rm fm^{-3}}$. Observational constraints from various pulsars, such as PSR J0740+6620, PSR J0614-3329, and GW measurement (e.g., GW170817) constraints are included to maintain the consistency between the DM-admixed NS results and astrophysical data. The value of $\beta$ is fixed at 1, 2, and 4 in the panels (a), b and c respectively. From the figure, it is observed that the inclusion of DM reduced both the stable maximum-mass and the radius of the NS. This behavior becomes more pronounced with the increase of $\alpha M_\chi$, since higher values correspond to higher DM concentration. Because the DM does not significantly contribute to the internal pressure, its gravitational presence enhances the compression of the NS to a more compact structure. 

In panel (a) ($\beta=1$), the mass-radius profile is consistent with both pulsar and GW constraints, including the $2 \ M_\odot$ limit, for certain values of $\alpha M_\chi$. However, for higher values of $\alpha M_\chi$, the profiles deviate from observational bounds. In panels (b) and (c), corresponding to higher values of $\beta=2$ and $4$, the mass-radius profile shifts more significantly, deviating from astrophysical constraints, indicating that a steeper DM profile (larger $\beta$) leads to a greater degree of core compression due to more centrally concentrated DM. In summary, high values of both $\alpha M_\chi$ and $\beta$ jointly cause the mass-radius profile to deviate from observational limits. This indicates that intensely concentrated DM distributions might produce excessively compact structures, which would be at conflict with the currently measured NS masses and radii.

Another important parameter which extract the crucial information about the contribution of DM to the NS structure and observables is the DM fraction inside the NS ($f_\chi$). This parameter explain that how the presence of DM affects the internal structure and total gravitational mass of the DM admixed NS. The DM fraction can be calculated by using the following formula,
\begin{eqnarray}
    f_\chi = \frac{M_{\rm DM}}{M}.
    \label{eq:DM_frac}
\end{eqnarray}

\begin{figure}
    \centering
    \includegraphics[width=0.5\textwidth]{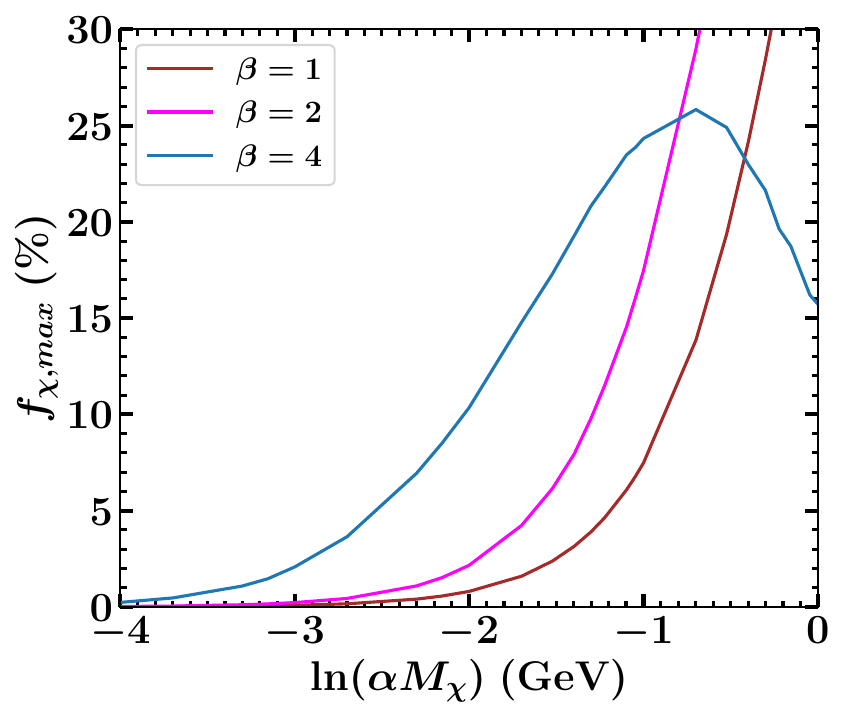}
    \caption{The maximum DM mass fraction is plotted with $\alpha M_\chi$ at different $\beta$ for Hornick3 EOS \cite{Hornick_2018}}.
    \label{fig:fx_alpMx}
\end{figure}

Here $M_{\rm DM}=\int_{0}^{R} 4\pi r^2 {\cal{E}}_{\rm DM} dr$ is the mass calculated using DM energy density and $M$ is the total mass of the DM admixed NS. In Fig. \ref{fig:fx_alpMx}, the DM fraction corresponding to the maximum-mass configuration, $f_{\chi , max}$, is shown as a function of $\alpha M_\chi$ for $\beta=1,2,$ and $4$. The results indicate that $f_{\chi , max}$ generally increases with $\alpha M_\chi$, reflecting the stronger contribution of heavier DM particles and larger scaling factors. For a fixed $\alpha M_\chi$, higher $\beta$ values yield larger $f_{\chi , max}$, highlighting the role of the steepness parameter in amplifying DM effects. However, for $\beta=4$, $f_{\chi , max}$ decreases beyond a critical $\alpha M_\chi$, showing that an excessively steep profile suppresses the DM contribution. This parameter space for $\beta=4$ diverges significantly from astrophysical constraints, consistent with earlier findings. This analysis shows that the $f_{\chi , max}$ provides a direct connection between microscopic DM parameters and macroscopic NS observables, offering a sensitive probe of how different DM properties influence NS structure. 

\subsection{Dark matter effects on Quasinormal modes}
\begin{figure}
   \centering
   \includegraphics[width = 0.5\textwidth]{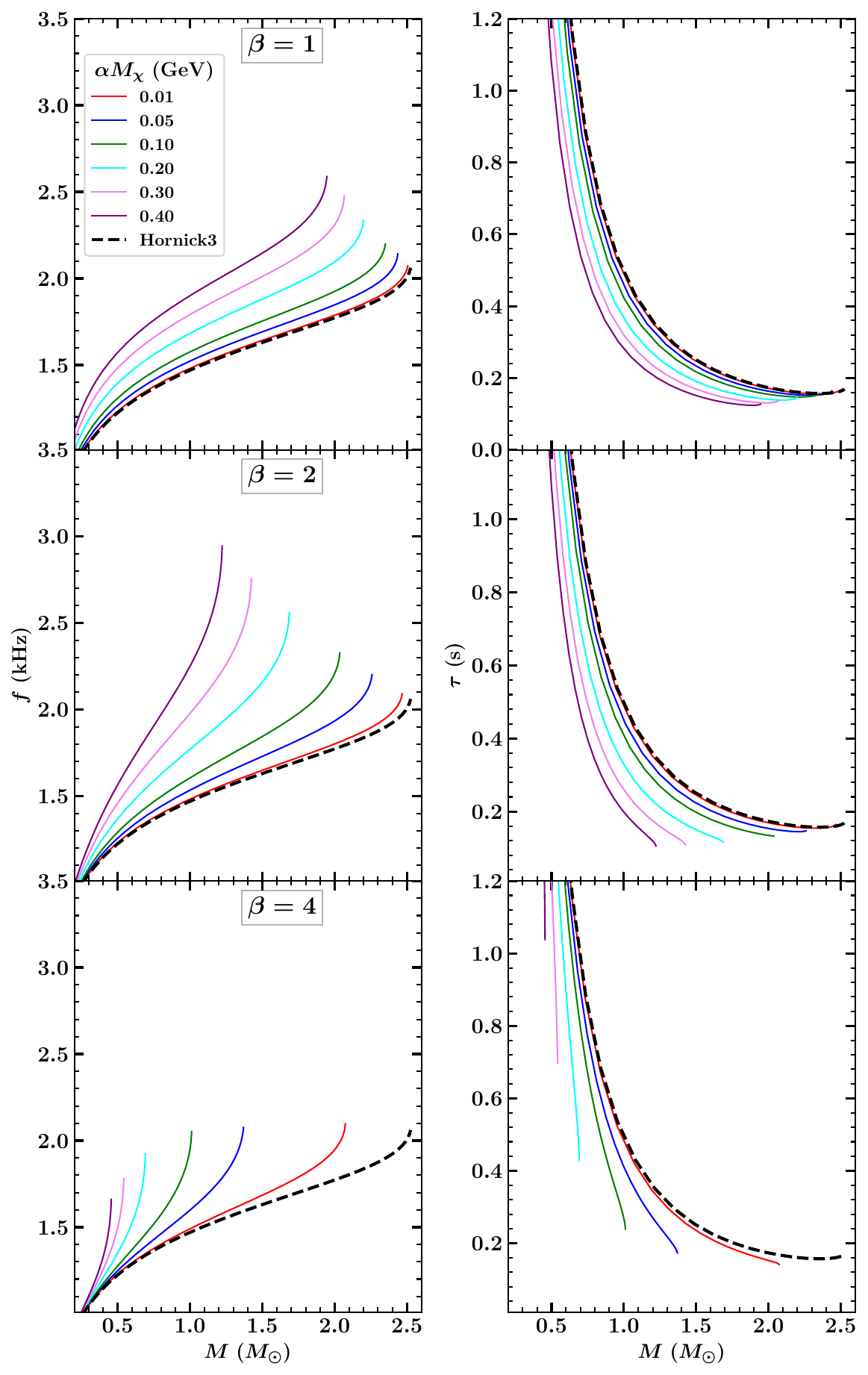}
   \caption{The $f$-mode frequency (left panel) and $\tau$ (right panel) are plotted with mass of the DM-admixed NS for Hornick3 EOS \cite{Hornick_2018}}.
    \label{fig:fd_m}
\end{figure}
The objective of this section is to investigate the relationship between the stellar mass, dimensionless tidal deformability, and the complex quasinormal mode frequencies for DM admixed NS. In Fig. \ref{fig:fd_m}, the $f$-mode frequency (left panel) and the corresponding $\tau$ (right panel) are shown as functions of mass. The effects of DM are incorporated through variations in both $\alpha M_\chi$ and $\beta$. For comparison, the pure hadronic case is also plotted with a black dashed line. When DM is included at $\beta=1$, the $f$-mode frequency shifts to higher values, and its magnitude continues to increase with increasing $\alpha M_\chi$. A similar trend is observed for larger values of $\beta$. Physically, higher values of $\beta$ and $\alpha M_\chi$ amplify the gravitational influence of the DM component, which in turn enhances the excitation of the $f$-mode frequency for a given stellar mass. For example, for a $1 \ M_\odot$ NS model with $\alpha M_\chi=0.05$, the calculated $f$-mode frequencies are approximately $1.52$ kHz, $1.53$ kHz, and $1.59$ kHz for $\beta=1, 2,$ and $4$, respectively. In contrast, the $\tau$ exhibits the opposite behavior to the $f$-mode frequency: it decreases with increasing stellar mass. This trend is anticipated since the $\tau$ is the inverse of the imaginary part of the complex eigen frequency. The presence of DM, introduced through variations in $\alpha M_\chi$ and $\beta$, further reduces the $\tau$ relative to the pure hadronic case. Numerical results support this behavior. For example, for a $1 \ M_\odot$ NS model with $\alpha M_\chi=0.05$, the calculated $\tau$ are $0.46$ s, $0.45$ s, and $0.41$ s for $\beta=1, 2,$ and $4$, respectively, thereby confirming the decreasing trend that is opposite to the increase observed in the $f$-mode frequency. 
\begin{figure}
   \centering
   \includegraphics[width = 0.5\textwidth]{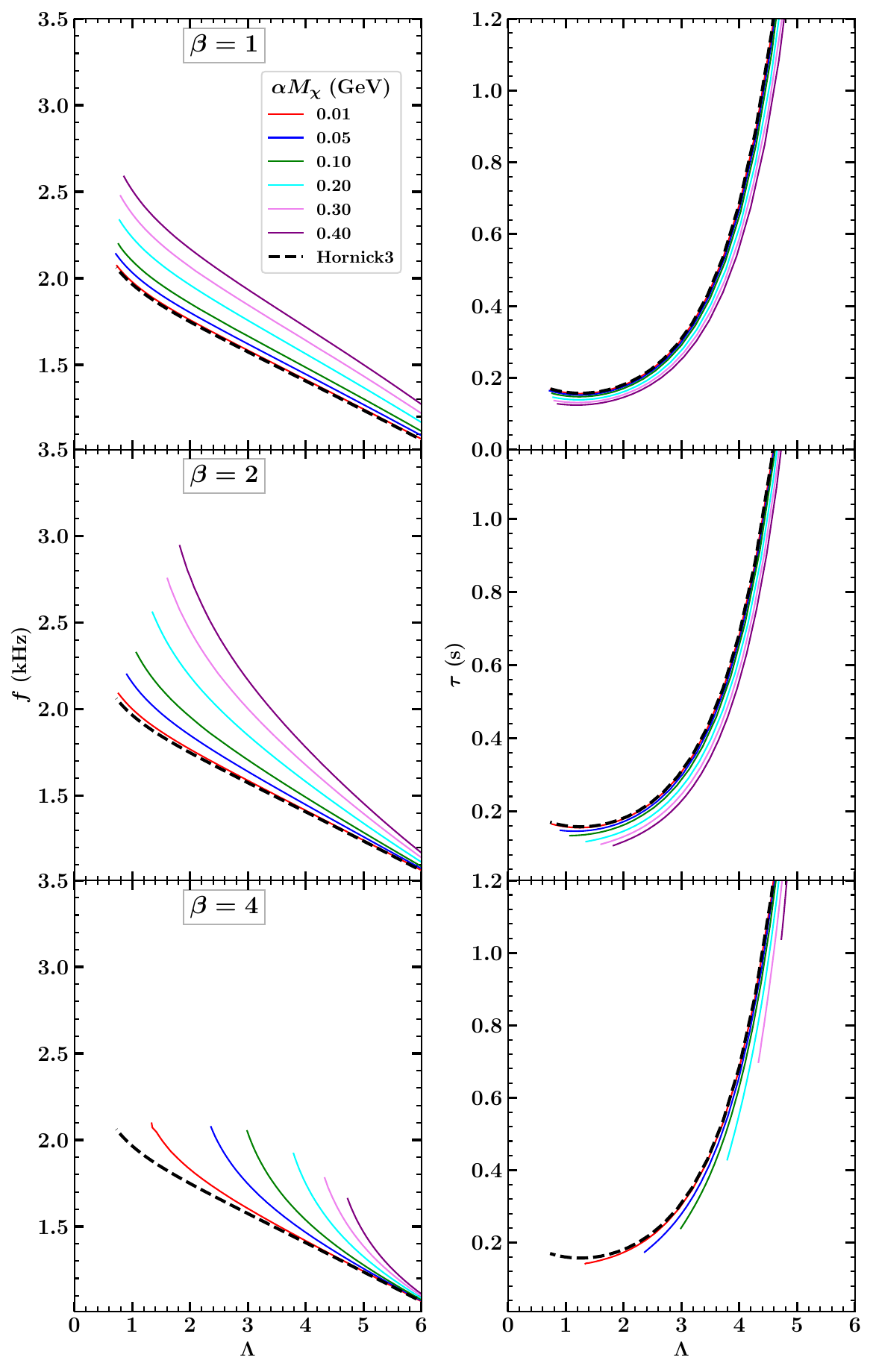}
   \caption{Same as Fig. \ref{fig:fd_m}, but with tidal deformability.}
    \label{fig:fd_l}
\end{figure}
In Fig. \ref{fig:fd_l}, the $f$-mode frequency (left panel) and the corresponding $\tau$ (right panel) are plotted as functions of the dimensionless tidal deformability $\Lambda$. In contrast to their dependence on mass, the $f$-mode frequency decreases with increasing $\Lambda$, while the $\tau$ increases. The effects of DM are also examined here: for a fixed value of $\Lambda$, the $f$-mode frequency increases in the presence of DM, whereas the $\tau$ decreases. A more detailed exploration of both the $f$-mode and the $\tau$, together with observational constraints from GW events, will be presented in a later section.

\begin{figure}
   \centering
   \includegraphics[width = 0.5\textwidth]{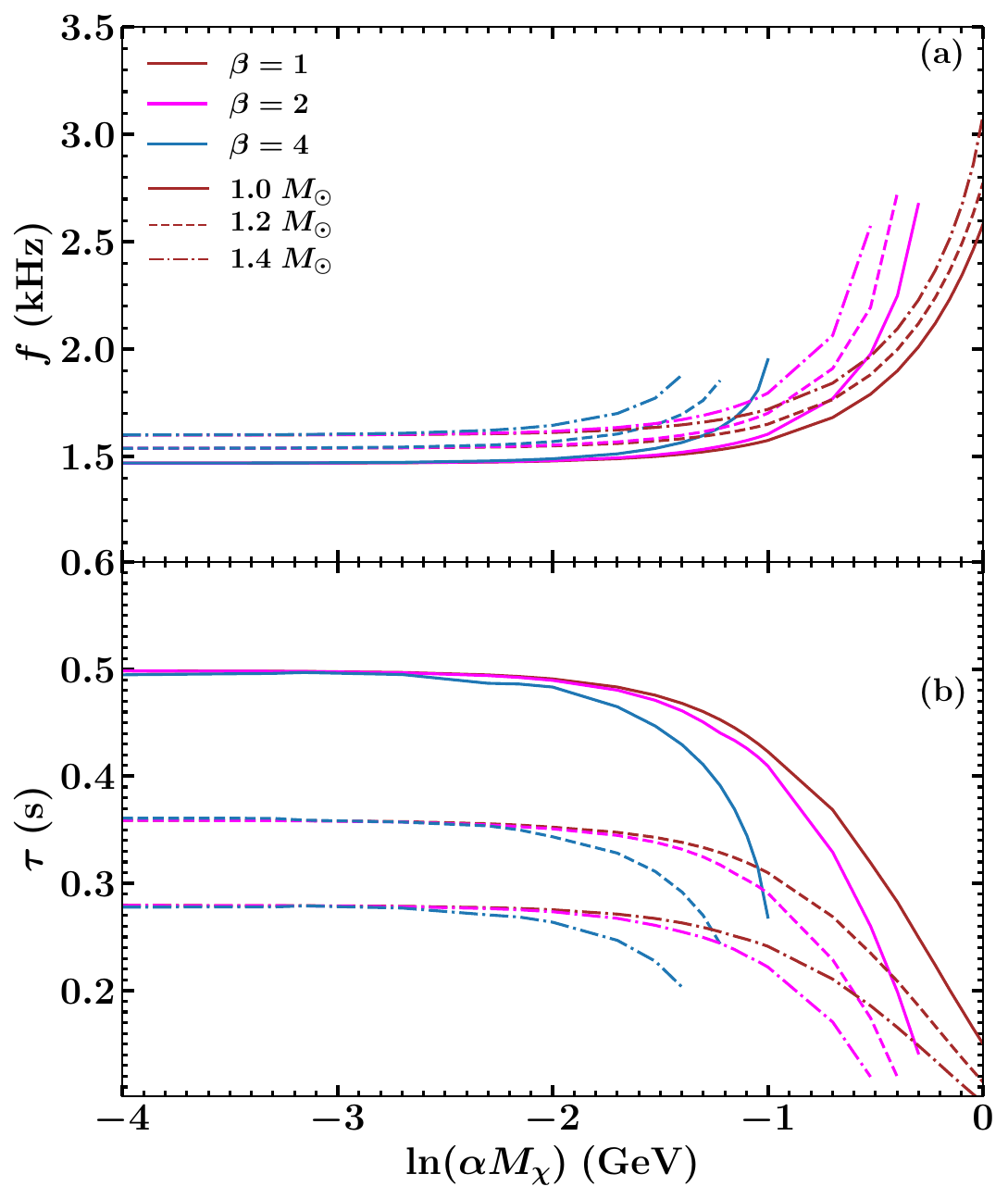}
   \caption{Panel (a): The $f$-mode frequency is shown for $1.0 \ M_\odot$, $1.2 \ M_\odot$, and $1.4 \ M_\odot$ DM-admixed NS models as a function of $\alpha M_\chi$ at fixed $\beta$ for Hornick3 EOS \cite{Hornick_2018}. Panel (b): Same as panel (a), but for the $\tau$.}
    \label{fig:fd}
\end{figure}
The effect of DM on both the $f$-mode frequency and the $\tau$ over a broader parameter space is explored in Fig. \ref{fig:fd}. To capture this dependence, $\alpha M_\chi$ is varied across a wide range for different values of $\beta$. Calculations are performed for DM admixed NS models with masses of $1.0 \ M_\odot$, $1.2 \ M_\odot$, and $1.4 \ M_\odot$. The high-mass model yields a higher $f$-mode frequency for a given $\alpha M_\chi$ compared to the lower-mass models, while the opposite behavior is observed for the $\tau$. At small $\alpha M_\chi$, the influence of $\beta$ is negligible across all models. However, as $\alpha M_\chi$ increases, the effect of DM becomes increasingly pronounced for different $\beta$. Notably, a sharp rise in the $f$-mode frequency, accompanied by a corresponding drop in the $\tau$, appears beyond a critical $\alpha M_\chi$, marking the regime where the DM strongly influences the stellar structure.

\begin{figure}
   \centering
   \includegraphics[width = 0.5\textwidth]{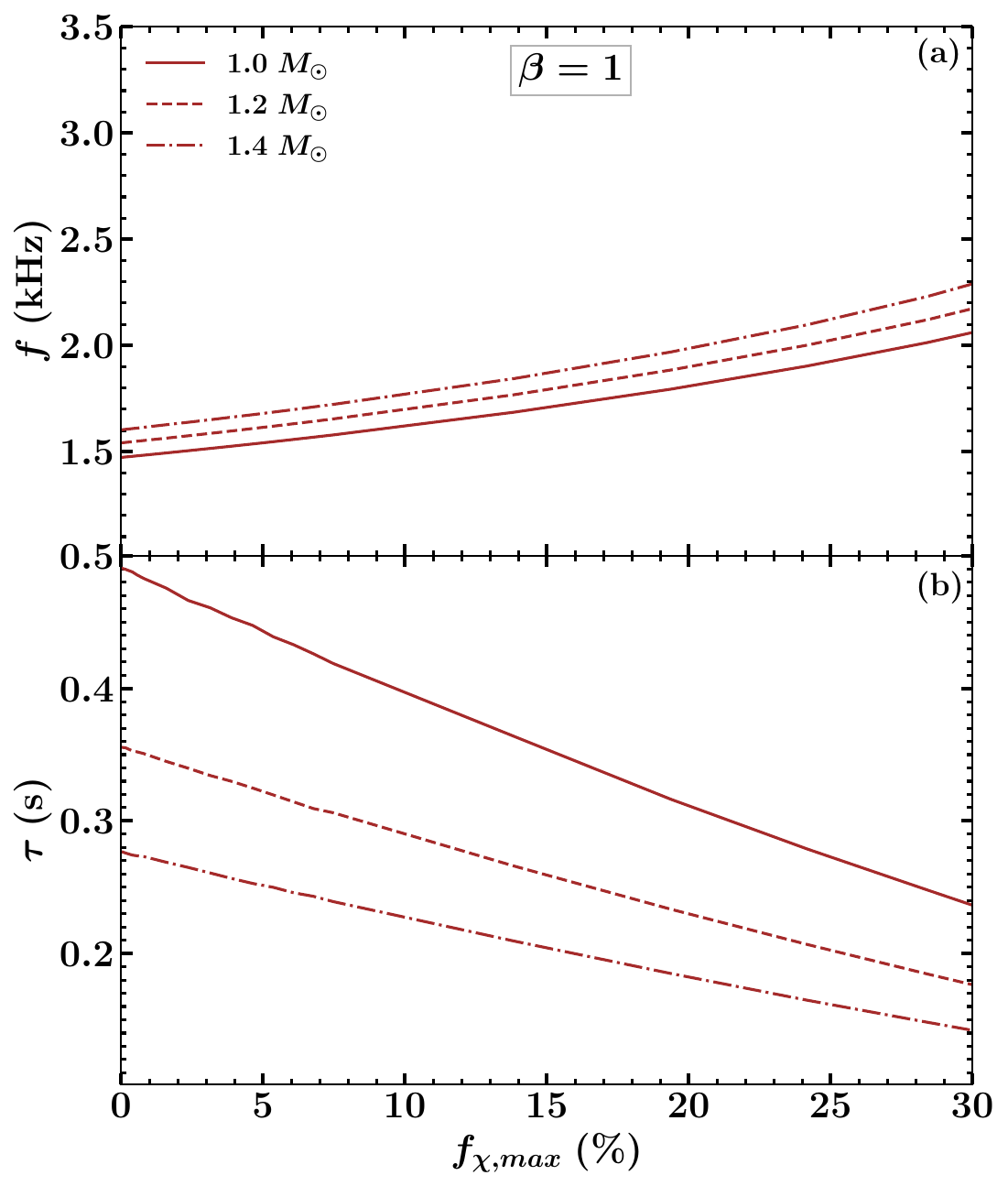}
   \caption{Panel (a): The $f$-mode frequency is shown for $1.0 \ M_\odot$, $1.2 \ M_\odot$, and $1.4 \ M_\odot$ DM-admixed NS models as a function of $f_{\chi , max}$ at $\beta=1$ for Hornick3 EOS \cite{Hornick_2018}. Panel (b): Same as panel (a), but for the $\tau$.}
    \label{fig:fd_fx}
\end{figure}
The effective controlling factor $\alpha M_\chi$ plays a key role in determining the dynamical properties of the star and simultaneously governs the $f_{\chi , max}$ within the NS. To explore how the $f_{\chi , max}$ influences these dynamical characteristics, Fig. \ref{fig:fd_fx} illustrates the variation of the complex QNM frequency with $f_\chi$. Similar to Fig. \ref{fig:fd}, the calculations are performed for three NS models: $1 \ M_\odot$, $1.2 \ M_\odot$, and $1.4 \ M_\odot$. It is found that as the $f_{\chi , max}$ within the star increases, the $f$-mode frequency (panel-a) also increases. Moreover, at a fixed $f_{\chi , max}$, higher-mass models exhibit larger $f$-mode frequencies compared to lower-mass ones. Conversely, the $\tau$ (panel-b) shows the opposite behavior, decreasing with increasing $f_{\chi , max}$ which indicates that the presence of DM enhances the rate of gravitational radiation damping, particularly in more massive and DM-rich NSs.

\begin{figure*}
   \centering
   \includegraphics[width = 0.5\textwidth]{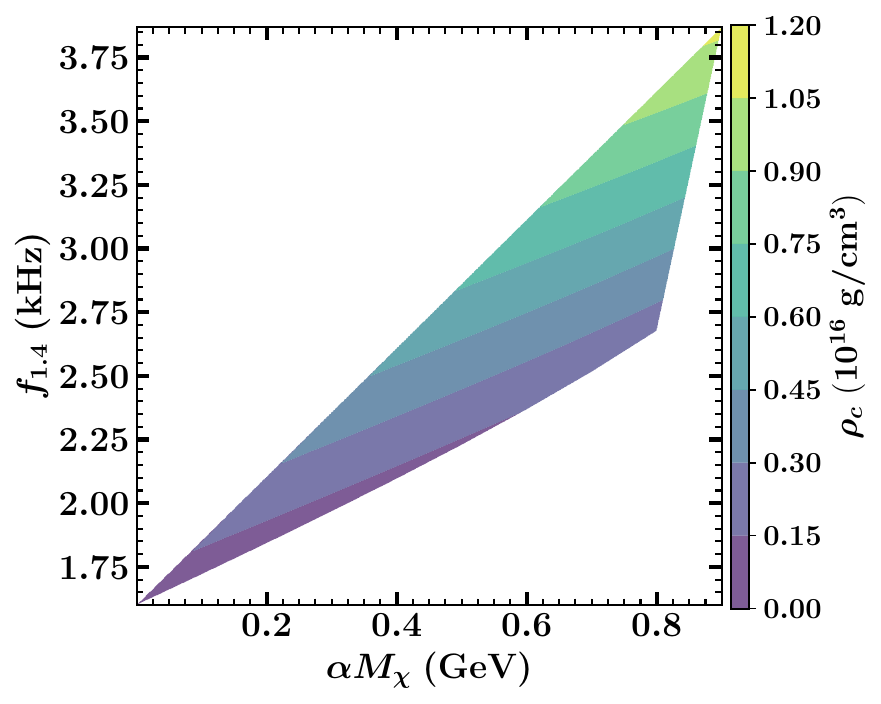}
   \includegraphics[width = 0.5\textwidth]{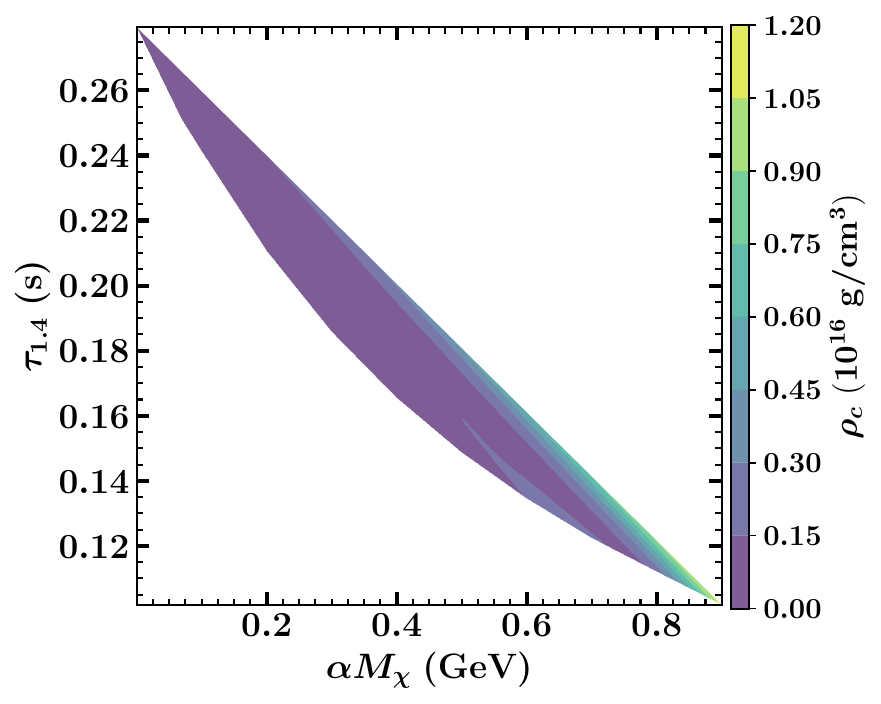}
   \caption{{\it Left panel:} The variation of the $f$-mode frequency for the canonical NS model as a function of the $\alpha M_\chi$. The color bar indicates the corresponding central energy densities. {\it Right panel:} Same as the left panel, but for the $\tau$. For both panel, we fix $\beta=1$ for the Hornick3 EOS.}
    \label{fig:contour}
\end{figure*}
In Fig.~\ref{fig:contour}, the dependence of the canonical $f$-mode frequency and the corresponding $\tau$ with respect to the parameter $\alpha M_\chi$ is illustrated. The color map represents the change in central energy density. The inclusion of DM in the NS softens the overall EOS, as DM contributes effectively to the energy density while providing only a minimal contribution to the pressure. Consequently, DM-admixed NSs exhibit higher central energy densities for increasing DM content. As shown in the left panel, the $f$-mode frequency shifts toward higher values with increasing $\alpha M_\chi$, corresponding to higher central energy densities. This upward frequency shift, arising from the additional DM degrees of freedom, could potentially enhance the detectability of these modes in future GW observations. In contrast, the $\tau$ shows the opposite behavior, as discussed earlier. As depicted in the right panel, the $\tau$ decreases with increasing DM content and central energy density, implying that DM-rich NSs oscillate with shorter lived modes compared to their purely hadronic counterparts.

A similar study of DM effects on NS oscillation was recently presented in Ref. \cite{Husain_2026}, in which a self-interacting DM model motivated by neutron decay anomaly was assumed, $n\longrightarrow \chi + \phi$. Keeping the self-interaction strength ($G$) as free parameter, the effects of the DM is incorporated to calculate the NS properties. In contrast, the current study assumes gravitationally bound fermionic DM  that may be accumulated in NS through astrophysical process. The DM distribution is described a phenomenological density profile characterized by the parameters $\alpha M_\chi$ and $\beta$. The two approaches therefore represent different physical origins and distributions of DM within NSs. From the perspective of NS asteroseismology, both studies investigate the impact of DM on the $f$-mode oscillations and the associated GW damping times. In the neutron decay scenario, the resulting $f$-mode frequencies and damping times were shown to follow approximate URs even in the presence of DM. The present work explores analogous relations for gravitationally bound DM-admixed NSs and examines how the presence of a centrally concentrated DM modifies the correlations between oscillation properties and macroscopic NS parameters. In particular, we analyze the behavior of the $f$-mode frequency and damping time in relation to stellar compactness and tidal deformability, thereby extending the UR framework to scenarios where DM alters the internal density distribution. This comparison highlights that while the detailed quantitative shifts depend on the assumed DM microphysics and distribution, the overall structure of the asteroseismic relations remains robust, suggesting that such relations may provide a useful probe of DM effects in NSs.

\begin{table*}[htbp]
    \centering
    \caption{The canonical value of NS properties with their corresponding $\alpha M_\chi$ values at different $\beta$ is shown for Hornick3 EOS.}
    \setlength{\tabcolsep}{5mm}
    \renewcommand{\arraystretch}{1.1}
    \begin{tabular}{|c|c|c|c|c|c|}
    \hline
     $\beta$ & $\alpha M_\chi$ & $R_{1.4}$ & $\Lambda_{1.4}$ & $f_{1.4}$ & $\tau_{1.4}$ \\
    \hline
    \multirow{6}{*}{1}
    & 0.01 & 13.18 & 706.31 & 1.61 & 0.27 \\
    & 0.05 & 12.91 & 610.75 & 1.66 & 0.25 \\
    & 0.10 & 12.59 & 515.44 & 1.72 & 0.24 \\
    & 0.20 & 11.98 & 374.54 & 1.84 & 0.21 \\
    & 0.30 & 11.41 & 272.06 & 1.96 & 0.18 \\
    & 0.40 & 10.89 & 199.30 & 2.09 & 0.16 \\
    \hline
    \multirow{6}{*}{2}
    & 0.01 & 13.16 & 695.39 & 1.61 & 0.27 \\
    & 0.05 & 12.79 & 558.06 & 1.69 & 0.24 \\
    & 0.10 & 12.31 & 422.61 & 1.79 & 0.22 \\
    & 0.20 & 11.25 & 211.18 & 2.06 & 0.17 \\
    & 0.30 & 9.74 & 66.18 & 2.57 & 0.11 \\
    & 0.40 & - & - & - & - \\
    \hline
    \end{tabular}
    \label{tab:cano_hnk4}
\end{table*}

To further illustrate the impacts of the DM, we present canonical NS properties with varying the parameters $\alpha M_\chi$ and $\beta$ in Tab. \ref{tab:cano_hnk4}. The table shows that how the higher and values of $\alpha M_\chi$ at constant $\beta$ makes the DM concentration high and alter the NS properties effectively. In addition, larger values of $\beta$ result in a steeper and more centrally concentrated DM distribution, further influencing the NS properties.

\subsubsection{Correlation Study}
In the previous sections, the influence of DM on various NS observables and QNMs has been analyzed. In this section, a correlation study among the NS observables, DM parameters, and $f$-mode characteristics is performed. For the NS properties, both canonical ($R_{1.4}$, $\Lambda_{1.4}$) and maximum-mass configurations ($M_{\text{max}}$, $R_{\text{max}}$, $\Lambda_{\text{max}}$) are considered. Regarding the QNMs, the $f$-mode frequency and its corresponding $\tau$ are included for both the canonical ($f_{1.4}$, $\tau_{1.4}$) and maximum-mass ($f_{\text{max}}$, $\tau_{\text{max}}$) models. The DM parameters include $\alpha M_\chi$, and the DM fraction for both canonical ($f_{\chi,1.4}$) and maximum-mass ($f_{\chi,\text{max}}$) stars. The Pearson correlation coefficient ($r(x, y)$) is used to quantify the linear correlation between any two variables $(x, y)$, and is defined as \cite{Pearson_2008, Pradhan_2022},
\begin{eqnarray}
    r(x,y)=\frac{cov(x,y)}{\sqrt{cov(x,x)cov(y,y)}}.
\end{eqnarray}
Here $cov(x,y)=\frac{1}{N} \sum\limits_{i=1}^{N} (x_i-\bar{x})(y_i-\bar{y})$.
\begin{figure}
   \centering
   \includegraphics[width = 0.5\textwidth]{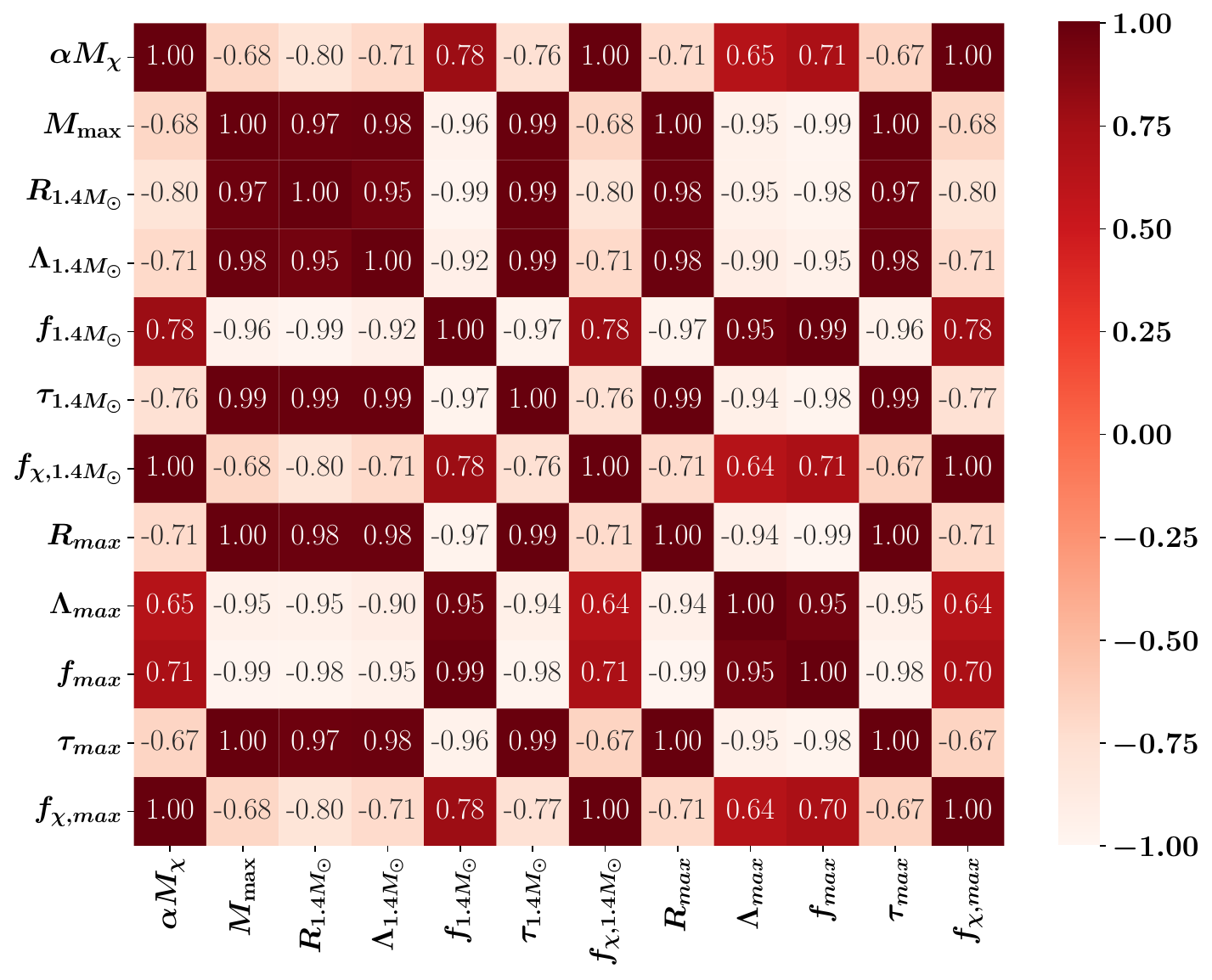}
   \caption{The heatmap represents the correlation matrix encompassing the correlation among NS observables, DM parameters, and $f$-mode characteristics. In this analysis, the steepness parameter is fixed at $\beta = 1$.}
    \label{fig:correlation}
\end{figure}
In this analysis, the DM effective controlling parameter is varied with fix steepness parameter ($\beta=1$) and the correlation matrix is shown in the Fig. \ref{fig:correlation}. A representative set of hadronic EOSs: QMC-RMF4 \cite{Alford_2022}, NITR-I \cite{pinku_ijmpe_2024}, Hornick1 \cite{Hornick_2018}, Hornick2 \cite{Hornick_2018}, Hornick3 \cite{Hornick_2018}, and Hornick4 \cite{Hornick_2018} are employed to construct this matrix, and the details of these EOSs are provided in Appendix \ref{Appendix-A}. Previously, in Ref. \cite{Pradhan_2022}, it was shown that for the nucleonic contribution as well as for both nucleonic and hyperonic contribution, the NS observables and also QNM characteristics are strongly correlated. Here for the DM-admixed NS, the following conclusions are drawn from the matrix, 
\begin{enumerate}[label=(\roman*)]
  \item A strong correlation was found among the NS observables even after the consideration DM. On the basis of Eq. \ref{eq:tidal}, it can be expected that the $\Lambda$ shows a strong correlation with $R$ ($0.95$ for canonical mass NS and $-0.94$ for maximum mass NS). Along with, frequency and damping time shows strong correlation.
  \item The Eq. \ref{eq:f_scaled} shows the correlation between the $f$-mode frequency and the radius, which is also seen in this matrix. The $f$-mode frequency is correlated strongly with the radius ($-0.99$ for canonical mass NS and $-0.99$ for maximum mass NS). Similarly, as expected from Eq. \ref{eq:damp_scaled}, a strong correlation is found between damping time and radius.
  \item The DM effective controlling parameter exhibits a strong positive correlation with both the canonical ($1.00$) and maximum ($1.00$) DM mass fractions. Moreover, the maximum DM mass fraction shows a perfect correlation with the canonical DM mass fraction.
\end{enumerate}

\subsection{Constraining dark matter parameter space}
\label{DM_par_space}
 
\begin{figure}[h]
   \centering
   \includegraphics[width=0.5\textwidth]{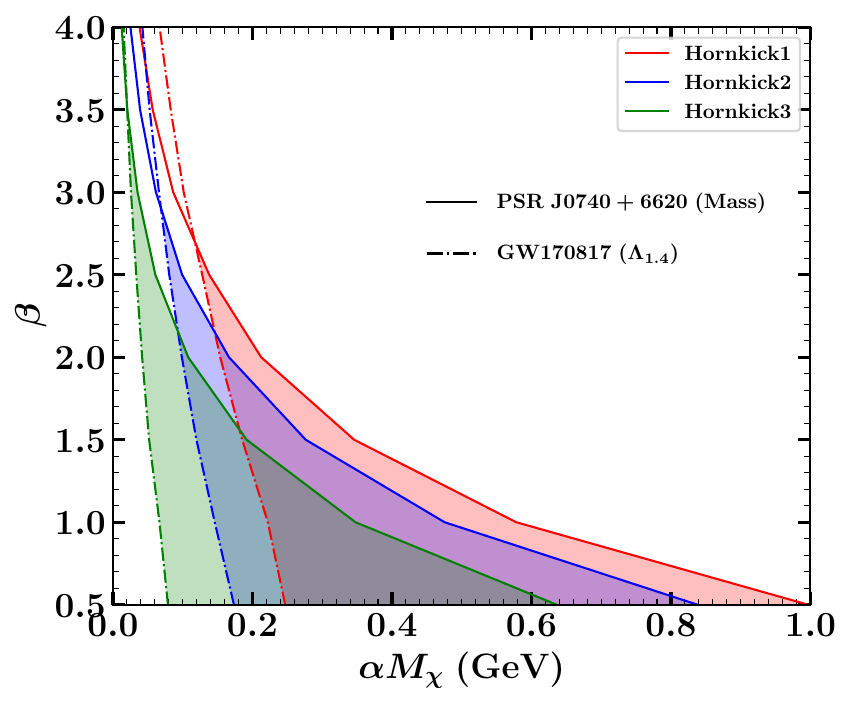}
   \caption{ The allowed range of $(\alpha M_\chi , \beta)$ parameter space is shown in different shaded region for each EOS; Hornick1 (red), Hornick2 (blue), Hornick3 (green). The shaded region is derived using observational constraints from  PSR J0740+6620 (maximum mass) and Gw170817 (canonical tidal deformability). }
   \label{fig:DM_Par}
\end{figure}

The physically plausible region of the $(\alpha M_\chi , \beta)$ parameter space is not assumed advance, but is instead identified using current observational constraints. In this section, we use observational constraints on neutron-star masses from NICER and on tidal deformability from the GW170817 event to identify the region of $(\alpha M_\chi , \beta)$ that remains compatible with existing data. The resulting allowed parameter space is shown in Fig. \ref{fig:DM_Par} and is obtained by imposing two key observational requirements:
\begin{itemize}
    \item Maximum mass constraint from PSR J0740+6620, $M_{\rm max}=2.08^{+0.07}_{-0.07} \ M_\odot$.
    \item Dimensionless tidal deformability constraint from GW170817, $\Lambda_{1.4}=190^{+390}_{-120}$.
\end{itemize}

For each EOS, the allowed parameter space is constructed by using the lower boundary of PSR J0740+6620 ($M_{\rm max}=2.01 \ M_\odot$) and upper boundary of GW170817 ($\Lambda_{1.4}=580$) for each EOS. To illustrate this procedure, we consider three relatively stiff EOSs in the present study: Hornick1 ($M_{max}=2.89 \ M_\odot$, $\Lambda_{1.4}=1145.81$) \cite{Hornick_2018}, Hornick2 ($M_{max}=2.72 \ M_\odot$ , $\Lambda_{1.4}=915.03$) \cite{Hornick_2018}, Hornick3 ($M_{max}=2.53 \ M_\odot$, $\Lambda_{1.4}=726.80$) \cite{Hornick_2018}. The detail parameter set of these EOSs are shown in Appendix \ref{Appendix-A} and can be found in original article Ref. \cite{Hornick_2018}.

As shown in Fig. \ref{fig:DM_Par}, increasing the value of $\beta$ leads to a progressively narrower allowed range of $\alpha M_\chi$. This trend reflects the fact that larger $\beta$ corresponds to a more strongly centrally concentrated DM distribution, which in turn produces a more compact stellar core and tighter stability requirements. In addition, stiffer EOSs are able to support a broader region of parameter space, owing to their greater ability to accommodate larger DM contributions while remaining consistent with the mass constraint from PSR J0740+6620. When tidal deformability constraints are taken into account, EOSs with larger intrinsic tidal deformabilities allow a wider range of $\alpha M_\chi$ values to remain below the GW170817 upper limit. Overall, this comparison highlights the non-trivial interplay between DM concentration, DM mass, and the underlying nuclear equation of state in determining viable DM–admixed NS configurations. It also emphasizes that the choice of nuclear EOS plays a crucial role not only in shaping NS properties, but also in defining the extent to which DM can be consistently incorporated within current observational bounds.

\begin{figure}[h]
   \centering
   \includegraphics[width=0.5\textwidth]{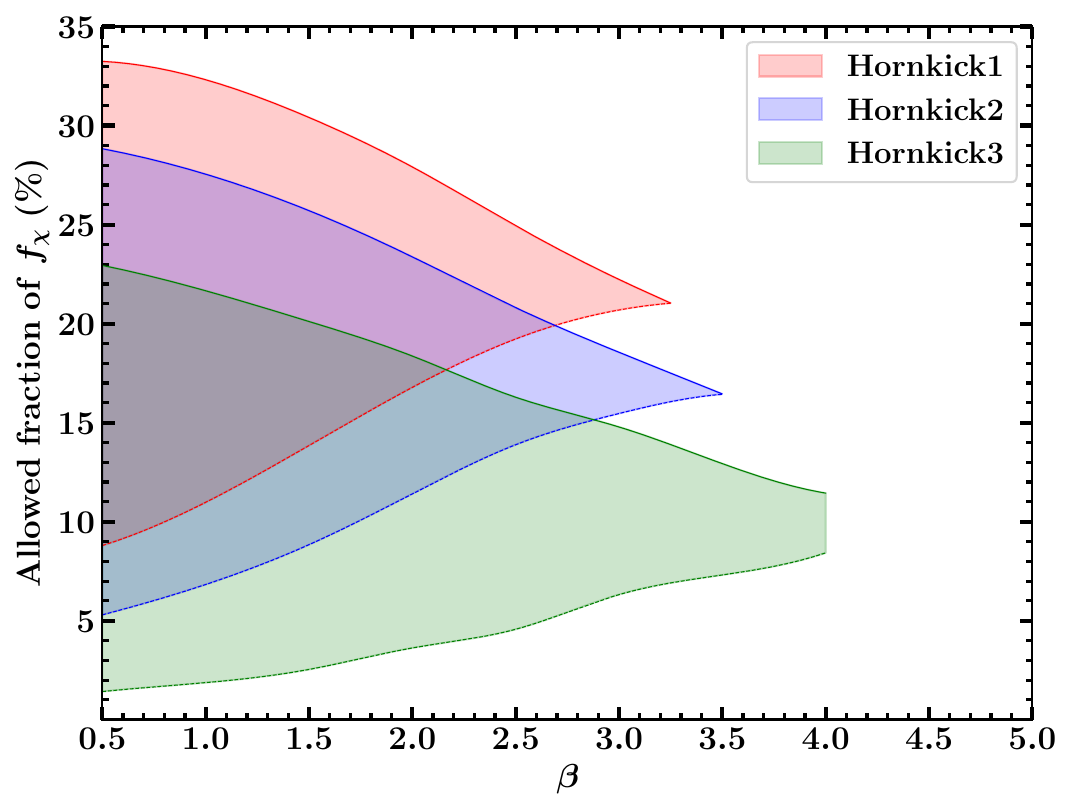}
   \caption{The allowed parameter regions of $f_\chi$ is shown for Hornick1, Hornick2, and Hornick3 EOSs using the observational constraint on mass with PSR J0740+6620 and tidal deformability with GW170817.}
   \label{fig:allowed_fx}
\end{figure}
In Fig. \ref{fig:allowed_fx}, we show the allowed DM mass fraction, $f_\chi$, as a function of $\beta$. The plausible regions of $f_\chi$ are obtained using the observational constraints from PSR J0740+6620 and GW170817. Considering different nuclear EOSs with varying maximum masses, the analysis further demonstrates how the allowed DM fraction depends on the stiffness of the underlying hadronic EOS. Among the considered EOSs, Hornick1 supports the largest allowed values of $f_\chi$ due to its relatively stiffer nature. For lower values of $\beta$, the maximum allowed DM fractions are approximately $33\%$ for Hornick1, $29\%$ for Hornick2, and $23\%$ for Hornick3, indicating that stiffer EOSs can accommodate comparatively larger DM contents while remaining consistent with current observational constraints. Another notable feature is that the allowed regions corresponding to different EOSs partially overlap. This suggests that, despite differences in the stiffness of the RMF models, the observational constraints impose common structural restrictions on the stellar configurations. As a result, even relatively softer EOSs can sustain moderate DM fractions for suitable choices of the DM parameters. Overall, the analysis indicates that the plausible range of $f_\chi$ is determined by the combined interplay between the DM parameters, the nuclear EOS, and the astrophysical observational constraints.

\subsection{Universal relations in DM-admixed NS asteroseismology}

The NS asteroseismology is the study of oscillation of NS which is a essential method for examining the internal composition of NS. The fundamental concept in NS asteroseismology aims to configure the angular frequency and GW damping timescale associated with an oscillation mode with the bulk characteristics of the NS, such as mass and radius. To investigate the URs, the present study employs Hornick3 EOS along with a representative set of hadronic EOSs: QMC-RMF4 \cite{Alford_2022}, NITR-I \cite{pinku_ijmpe_2024}, Hornick1 \cite{Hornick_2018}, Hornick2 \cite{Hornick_2018}, and Hornick4 \cite{Hornick_2018}. The Lagrangian density used to generate these EOSs, along with the corresponding parameter values, is provided in Appendix \ref{Appendix-A}.

The concept of NS asteroseismology was first introduced by Andersson and Kokkotas \cite{Andersson_1996, Andersson_1998}. In this framework, the $f$-mode oscillation is of particular interest, as its frequency scales approximately linearly with the average stellar density, while its damping time varies inversely with stellar compactness when normalized by the factor $M^3/R^4$. These dependencies can be expressed as,

\begin{eqnarray}
    f({\rm kHz}) = a_r + b_r\sqrt{\frac{M}{R^3}},
    \label{eq:f_scaled}
\end{eqnarray}

\begin{eqnarray}
    \frac{R^4}{M^3\tau_f} = a_i + b_i\frac{M}{R}.
    \label{eq:damp_scaled}
\end{eqnarray}

\begin{figure}
   \centering
   \includegraphics[width = 0.5\textwidth]{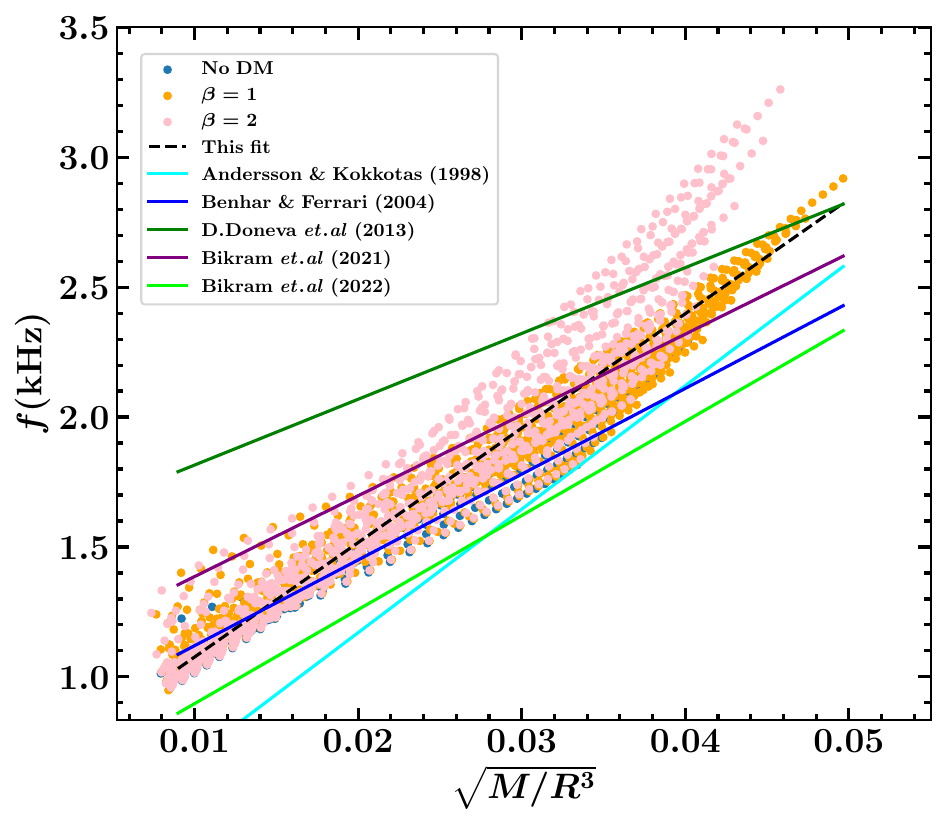}
   \caption{Variation of the $f$-mode frequency with average density. For comparison, fits from Andersson \& Kokkotas \cite{Andersson_1998}, Benhar \& Ferrari \cite{Benhar_2004}, Doneva {\it et al.} \cite{Doneva_2013}, and Bikram {\it et al.} \cite{Pradhan_2021, Pradhan_2022} are also shown.}
    \label{fig:fden}
\end{figure}

\begin{figure}
   \centering
   \includegraphics[width = 0.5\textwidth]{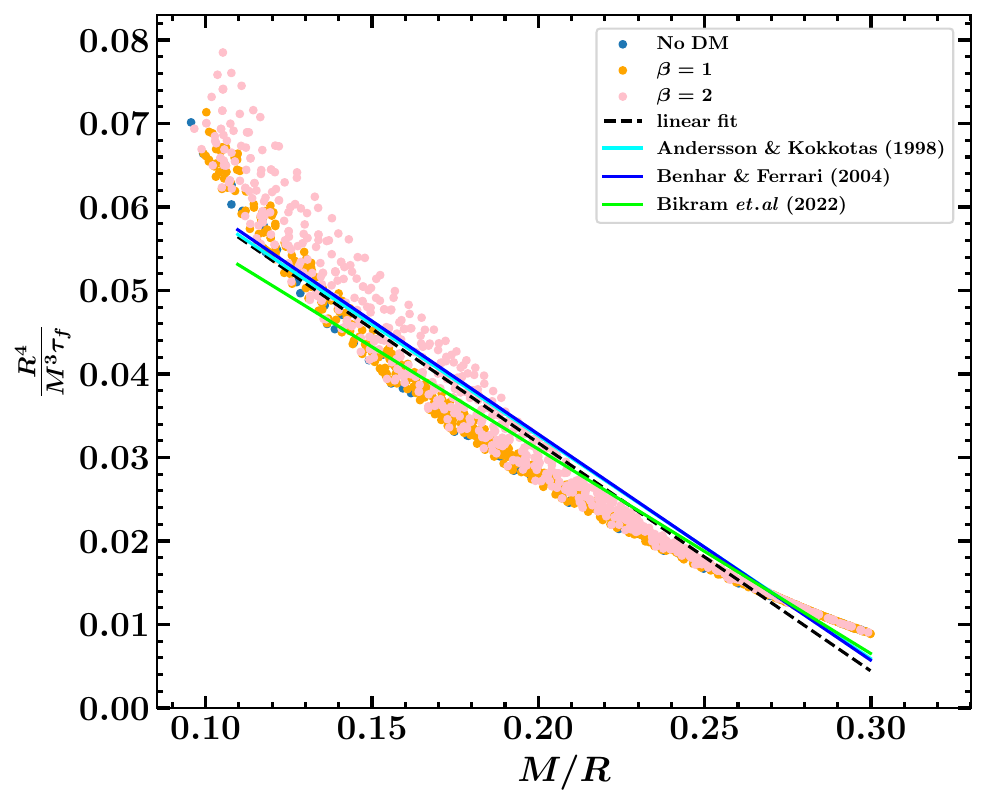}
   \caption{Variation of scaled damping time with compactness ($C=M/R$). Fits from Andersson \& Kokkotas \cite{Andersson_1998}, Benhar \& Ferrari \cite{Benhar_2004} and Bikram {\it et al.} \cite{Pradhan_2022} additionally displayed for comparison.}
    \label{fig:damp_scaled}
\end{figure}

Here, $a_r$, $b_r$, $a_i$, and $b_i$ denote the fitting coefficients obtained from the best fit to the data. This relation has been further explored in the literature, including extensions to exotic degrees of freedom such as quarks and hyperons \cite{Benhar_2004, Salcedo_2014}. The effects of rotation on these fits were later analyzed by Doneva {\it et al.} \cite{Doneva_2013}. In Ref. \cite{Pradhan_2021}, Bikram and Chatterjee examined this relation for NSs with hyperonic cores within the Cowling approximation, and subsequently extended their study to a full general relativistic framework in Ref. \cite{Pradhan_2022}. In the present work, the adopted DM model fully accounts for the gravitational impact on the DM density distribution. Therefore, its effect on the above fits requires careful examination. To this end, we investigate the $f$-mode scaling relations by considering both nucleonic matter and DM, up to relatively high DM concentrations varying both $\alpha M_\chi$ and $\beta$. The $f$-mode frequency, scaled with the average density, is shown in Fig. \ref{fig:fden}, and the best-fit coefficients $a_r$ and $b_r$ are summarized in Table \ref{tab:f_avg-den}. The present analysis yields $a_r = 0.686 \ {\rm kHz}$ and $b_r = 40.535 \ {\rm kHz \ km}$. These values are very different from previous results, reflecting the sensitivity of the $f$-mode frequency to the DM distribution. Nevertheless, despite the numerical deviations induced by DM, the overall scaling with the average density remains consistent. This reinforces the robustness of the $f$-mode as a global oscillation mode, and highlights its potential utility for astrophysical observations. Similarly, the normalized $\tau$ as a function of stellar compactness is displayed in Fig. \ref{fig:damp_scaled}, with the corresponding best-fit coefficients $a_i$ and $b_i$ listed in Table \ref{tab:d-scaled_C}. The present work yields $a_i = 0.086$ and $b_i = -0.273$.

\begin{table}[h]
    \caption{The fitting coefficients $a_r$ and $b_r$ for the $f$-mode frequency obtained in the present work, together with results from previous studies, are summarized. These coefficients are determined using the asteroseismology relation given in Eq. \ref{eq:f_scaled}.}
    \centering
    \renewcommand{\tabcolsep}{0.2cm}
    \renewcommand{\arraystretch}{1.3}
    \scalebox{1.0}{
    \begin{tabular}{lcc}
    \hline \hline
    Fittings & $a_r$ (kHz) & $b_r$ (kHz$\times$km) \\
    \hline
    Andersson \& Kokkotas (1998) \cite{Andersson_1998}     & 0.220 & 47.510 \\
    Benhar \& Ferrari (2004) \cite{Benhar_2004}            & 0.790 & 33.000 \\
    D. Doneva {\it et. al} (2013) \cite{Doneva_2013}       & 1.562 & 25.320 \\
    Bikram {\it et. al} (2021) \cite{Pradhan_2021}         & 1.075 & 31.100 \\
    Bikram {\it et. al} (2022) \cite{Pradhan_2022}         & 0.535 & 36.200 \\
    This work                                              & 0.636 & 44.038 \\ 
    \hline \hline
    \end{tabular}}
    \label{tab:f_avg-den}
\end{table}

\begin{table}[h]
    \caption{The fitting coefficients $a_i$ and $b_i$ for the damping time obtained in the present work, together with results from previous studies, are summarized. These coefficients are determined using the asteroseismology relation given in Eq. \ref{eq:damp_scaled}.}
    \centering
    \renewcommand{\tabcolsep}{0.35cm}
    \renewcommand{\arraystretch}{1.3}
    \scalebox{1.0}{
    \begin{tabular}{lcc}
    \hline \hline
    Fittings & $a_i$ &  $b_i$\\
    \hline
    Andersson \& Kokkotas (1998) \cite{Andersson_1998}      & 0.086 & $-$0.267 \\
    Benhar \& Ferrari  (2004) \cite{Benhar_2004}            & 0.087 & $-$0.271 \\
    Bikram {\it et. al} (2022) \cite{Pradhan_2022}          & 0.080 & $-$0.245 \\
    This work                                               & 0.086 & $-$0.273 \\ 
    \hline \hline
    \end{tabular}}
    \label{tab:d-scaled_C}
\end{table}

As discussed above, the fitting relations in Eq. \ref{eq:f_scaled} and \ref{eq:damp_scaled} demonstrate particular model reliance. Alongside these, however, there exist URs that remain largely insensitive to the details of the EOSs. Such relations are found to be robust across a wide variety of microphysical scenarios, including those involving exotic constituents. Because of this near independence, URs provide a more reliable approach for extracting NS properties from QNM observations, even when the microscopic composition of dense matter is uncertain. A variety of URs have been proposed to estimate the global properties of NSs \cite{Yagi_2013, Yagi_2015, Chirenti_2015, Staykov_2016, Breu_2016, Landry_2018, Kumar_2019, Jiang_2020, Sotani_2021, Zhao_2022, Pradhan_2022, sailesh_2024, Shirke_2024, pinku_mnras_2023}. Although the physical origin of these relations is not yet fully understood and remains a subject of ongoing research, in this study I adopt a more practical perspective, regarding URs as robust trends that naturally emerge in compact stars. From this viewpoint, URs provide a useful framework to constrain phenomenological quantities that are either difficult to measure directly or exhibit degeneracies in astrophysical observations. The present analysis focuses on URs associated with the $f$- mode of NS oscillations. Among the spectrum of oscillation modes, the $f$-mode is particularly significant because it couples strongly to both GWs and tidal excitations, making it one of the most promising candidates for detection in binary NS mergers \cite{Chakravarti_2020}. To examine this, I construct URs using the scaled (normalized) $f$-mode frequency, $\bar{\omega}=\omega M$, in terms of the dimensionless the stellar compactness $C=M/R$ and tidal deformability $\Lambda$.

\subsubsection{$\bar{\omega}-C$ relation}
The UR between the compactness ($C$) and the $f$-mode was originally established by Andersson and Kokkotas \cite{Andersson_1998}. Tsui \cite{Tsui_2005} and Lioutas \cite{Lioutas_2021} then expanded on this study using a wider range of EOSs to investigate GW asteroseismology. Here, I use the approximate formula derived from least-squares fitting to get the $\bar{\omega}-C$ relations:
 
\begin{equation}
\label{eq:f-C_fitting}
    \bar{\omega}(C) = \sum_{n=0}^{n=4} a_n (C)^n + i\sum_{n=0}^{n=6} a^{\prime}_{n} (C)^n \, .
\end{equation}

Here $f$-mode frequency is normalized with mass ($M$) by the relation $\bar{\omega}=\omega M$. The coefficients $a_n$ and $a_n^\prime$ shown in the Table \ref{tab:coeeficient_fd_c}. The residuals are computed with the formula, 
\begin{equation}
\label{eq:f_residual}
    \Delta_f = \frac{| \bar{\omega}_f - \bar{\omega}_f^{\rm fit} | }{\bar{\omega}_f^{\rm fit}} \quad \& \quad \Delta_\tau = \frac{| \bar{\omega}_\tau - \bar{\omega}_\tau^{\rm fit} | }{\bar{\omega}_\tau^{\rm fit}}  .
\end{equation}
Similarly, we can have the root mean square error (RMSE) for the fit using the formula, 
\begin{equation}
\label{eq:f_rmse}
    ({\rm RMSE})_i = \sqrt{\frac{1}{N}\sum \Delta_i ^2}, 
\end{equation}
where  $i= f, \tau$.

\begin{figure*}
   \centering
   \includegraphics[width=0.5\linewidth]{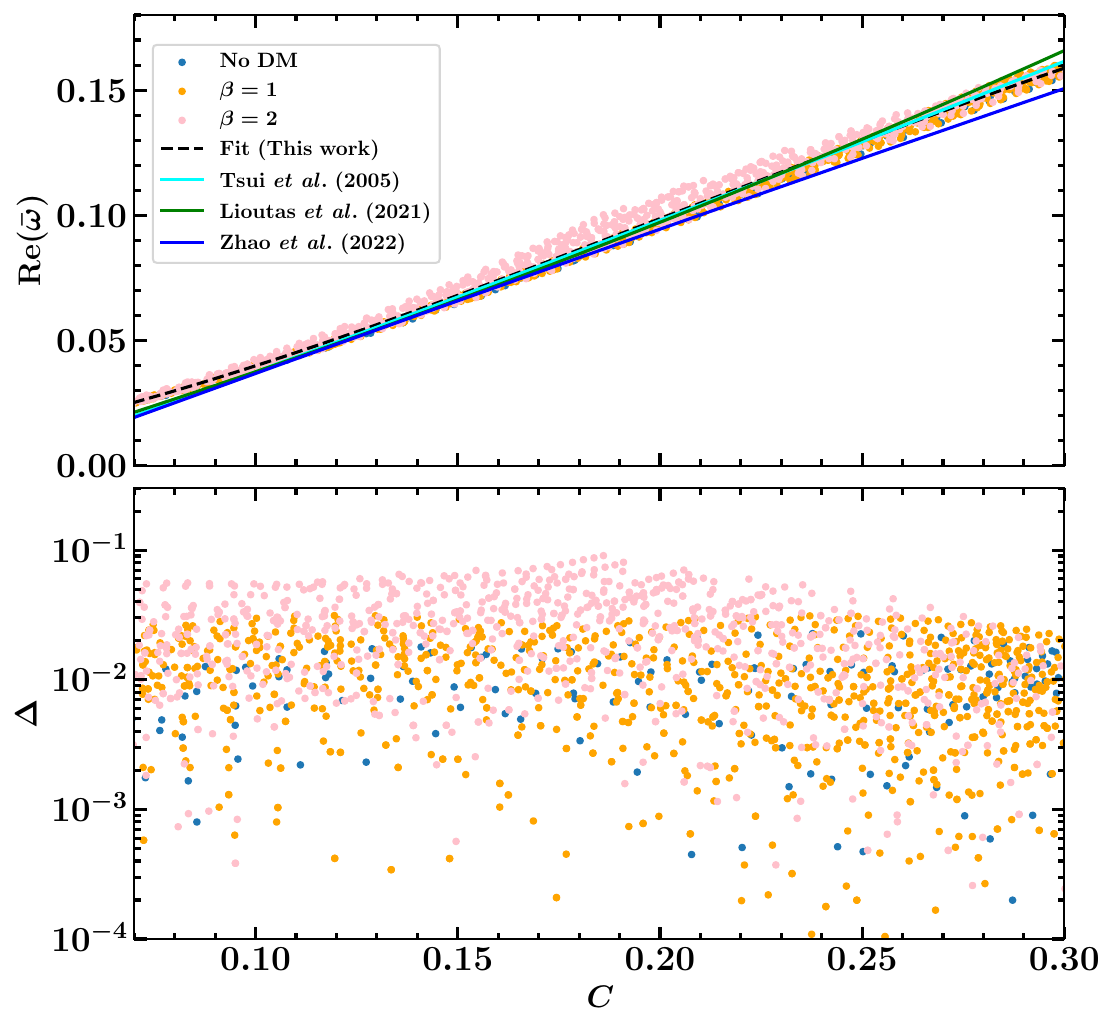}
   \includegraphics[width=0.5\linewidth]{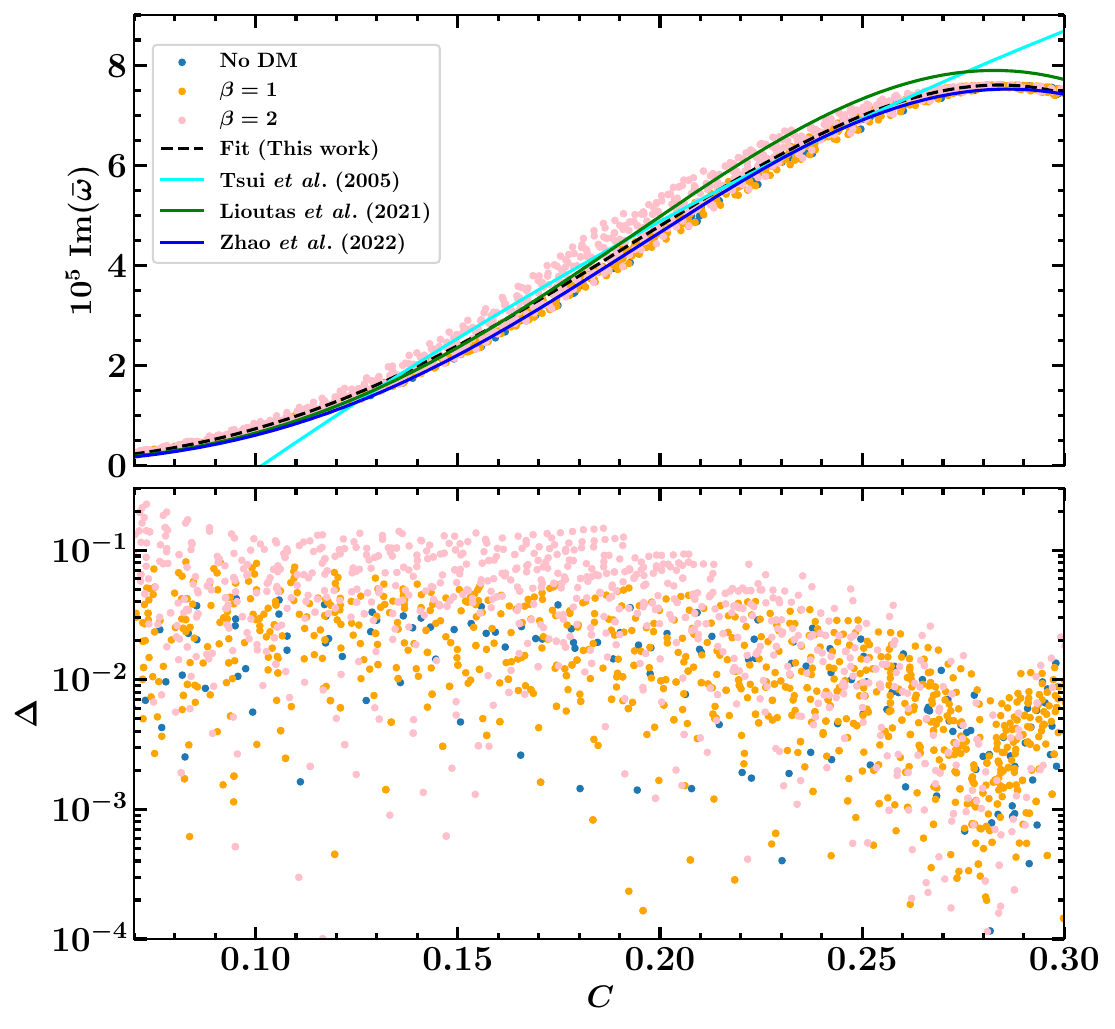}
   \caption{The UR between the scaled QNM frequency and stellar compactness is shown. The black dashed curve represents the least-squares fit obtained using Eq. \ref{eq:f-C_fitting}. For comparison, the fitting relations reported in Refs. Zhao {\it et al.} (2022) \cite{Zhao_2022}, Lioutas {\it et al.} (2020, 2021) \cite{Lioutas_2020, Lioutas_2021}, and Tsui {\it et al.} (2005) \cite{Tsui_2005} are also included. The universality for the real part ${\rm Re}(\bar{\omega})$ is displayed in the left panel, where as the universality of mass scaled damping time ${\rm Im}(\bar{\omega})$ is displayed in the right panel. The residuals corresponding to each fit shown in the bottom panels.}
   \label{fig:UR_Comp}
\end{figure*}

Utilizing Eq. \ref{eq:f-C_fitting}, the UR between the scaled QNM frequency and stellar compactness ($\bar{\omega}-C$) is shown in Fig. \ref{fig:UR_Comp}. The corresponding fitting coefficients are provided in Tab. \ref{tab:coeeficient_fd_c}, together with results from earlier studies.  The RMSE and maximum residual for real (imaginary) part are 0.028 (2.110) and 0.180 (17.69) respectively. Previous works have shown that when the frequency is scaled with stellar radius, noticeable deviations from universality appear \cite{Tsui_2005}. To address this, the present study adopts the more robust relation between the scaled QNM frequency and compactness \cite{Lioutas_2017, Pradhan_2022}. The present fit for DM-admixed NSs also follows the $\bar{\omega}-C$ relation closely without breaking universality, thereby reinforcing its robustness even in scenarios where DM is gravitationally captured and distributed non-uniformly.

\begin{table*}
    \caption{Fitting coefficients for the $\bar{\omega}$–$C$ UR shown using the approximate formula Eq. (\ref{eq:f-C_fitting}). The upper table lists the coefficients for the real part of the QNM frequency, while the lower table corresponds to the imaginary part.}
    \centering
    \renewcommand{\tabcolsep}{0.35cm}
    \renewcommand{\arraystretch}{1.3}
    \scalebox{0.8}{
    \begin{tabular}{lccccc}
    \hline \hline
    Fittings & $a_0(10^{-3})$ &  $a_1(10^{-1})$ &  $a_2(10^{-1})$ &  $a_3$ &  $a_4$\\
    \hline
    Tsui {\it et. al} (2005) \cite{Tsui_2005}     & $20.000$  & $5.600$  & $1.500$ &   $--$   & $--$ \\
    Lioutas {\it et. al} (2021) \cite{Lioutas_2021}  & $-13.220$ & $4.627$  & $4.466$ &   $--$   & $--$ \\
    Zhao {\it et. al} (2022) \cite{Zhao_2022}     & $-22.230$ & $5.982$  & $-0.733$ &   $--$   & $--$ \\
    This work                            & $1.697$   & $2.038$  & $21.923$ & $-4.285$ & $1.7686$\\ 
    \hline \hline
    \end{tabular}}\\ \vspace{0.2cm}
    \scalebox{0.8}{
    \begin{tabular}{lccccccc}
    \hline \hline
    Fittings & $a_0^\prime(10^{-6})$ &  $a_1^\prime(10^{-4})$ &  $a_2^\prime(10^{-3})$ &  
    $a_3^\prime(10^{-2})$ &  $a_4^\prime(10^{-1})$ &  $a_5^\prime(10^{-1})$ &  $a_6^\prime(10^{-1})$\\
    \hline
    Tsui {\it et. al} (2005) \cite{Tsui_2005}     & $-62.000$ & $6.700$  & $0.580$  & $--$    & $--$     & $--$     & $--$\\
    Lioutas {\it et. al} (2021) \cite{Lioutas_2021}  & $0.000$   & $0.000$  & $0.000$  & $0.000$ & $1.120$  & $-5.300$ & $6.280$\\
    Zhao {\it et. al} (2022) \cite{Zhao_2022}     & $0.000$   & $0.000$  & $0.000$  & $0.000$ & $1.048$  & $-4.971$ & $5.943$\\
    This work                            & $0.469$   & $0.030$  & $-0.718$ & $1.814$ & $-0.434$ & $--$  & $--$\\ 
    \hline \hline
    \end{tabular}}
    \label{tab:coeeficient_fd_c}
\end{table*}

\subsubsection{$\bar{\omega}-\Lambda$ relation}
The multi-messenger astronomy involving GW observation provides a new insight to constrain the NS properties theoretically utilizing the URs. The tidal deformability extracted from this observational can be used as a crucial parameter in URs to put constraints on NS properties. And the non-radial $f$-mode frequency which is a promising source of GWs can be calculated theoretically using the $f-\Lambda$ UR. In this work, i have calculated the $f-\Lambda$ relations and perform a least-square fit using the approximate formula:

\begin{equation}
\label{eq:f-Love_fitting}
    \bar{\omega} (\Lambda) = p(\log(\Lambda)) + i10^{q(\log(\Lambda))} ,
\end{equation}
where,
\[p(\log(\Lambda))= \sum_{n=0}^{n=7} b_n (\log(\Lambda))^n \quad \& \quad \]
\[q(\log(\Lambda))= \sum_{n=0}^{n=7} b^{\prime}_{n} (\log(\Lambda))^n 
.\]

\begin{figure*}
   \centering
   \includegraphics[width=0.5\linewidth]{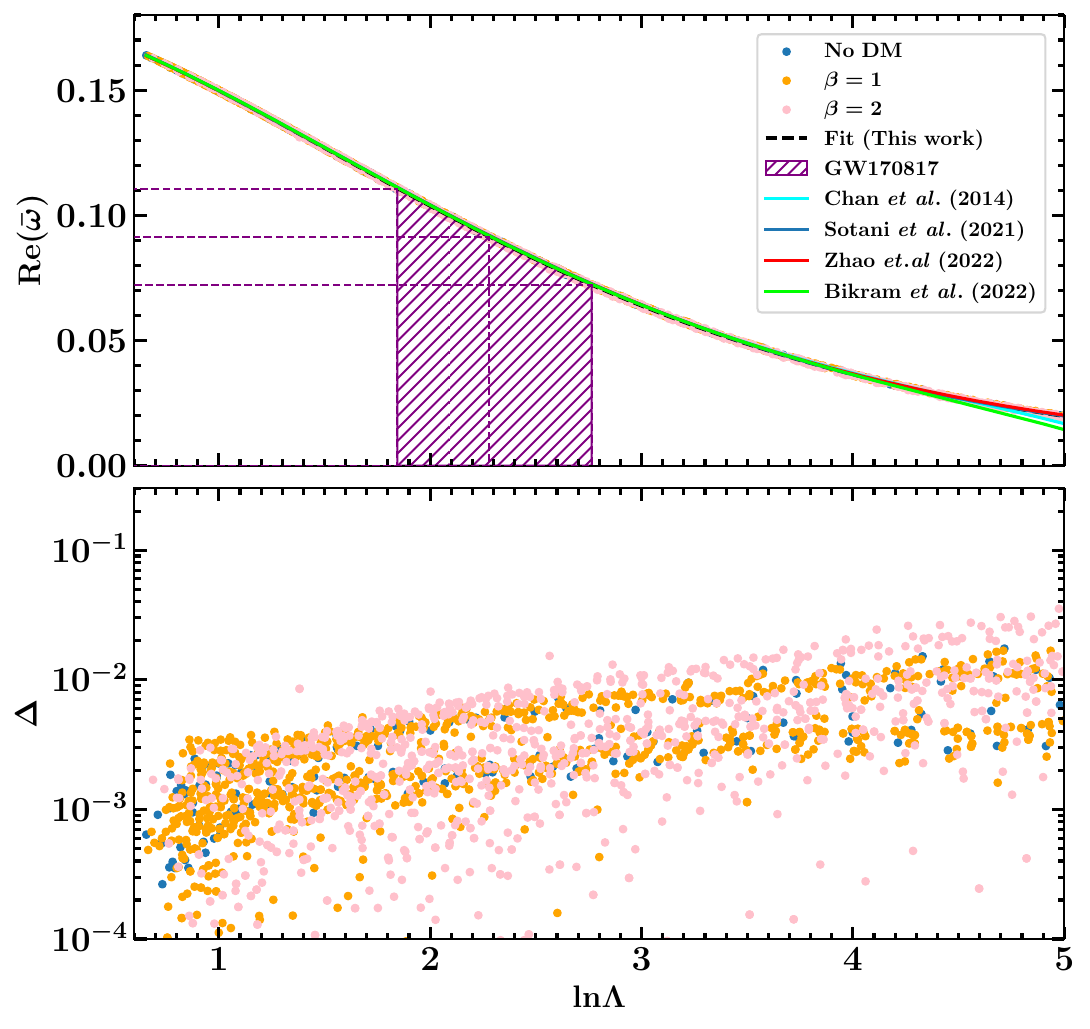}
   \includegraphics[width=0.5\linewidth]{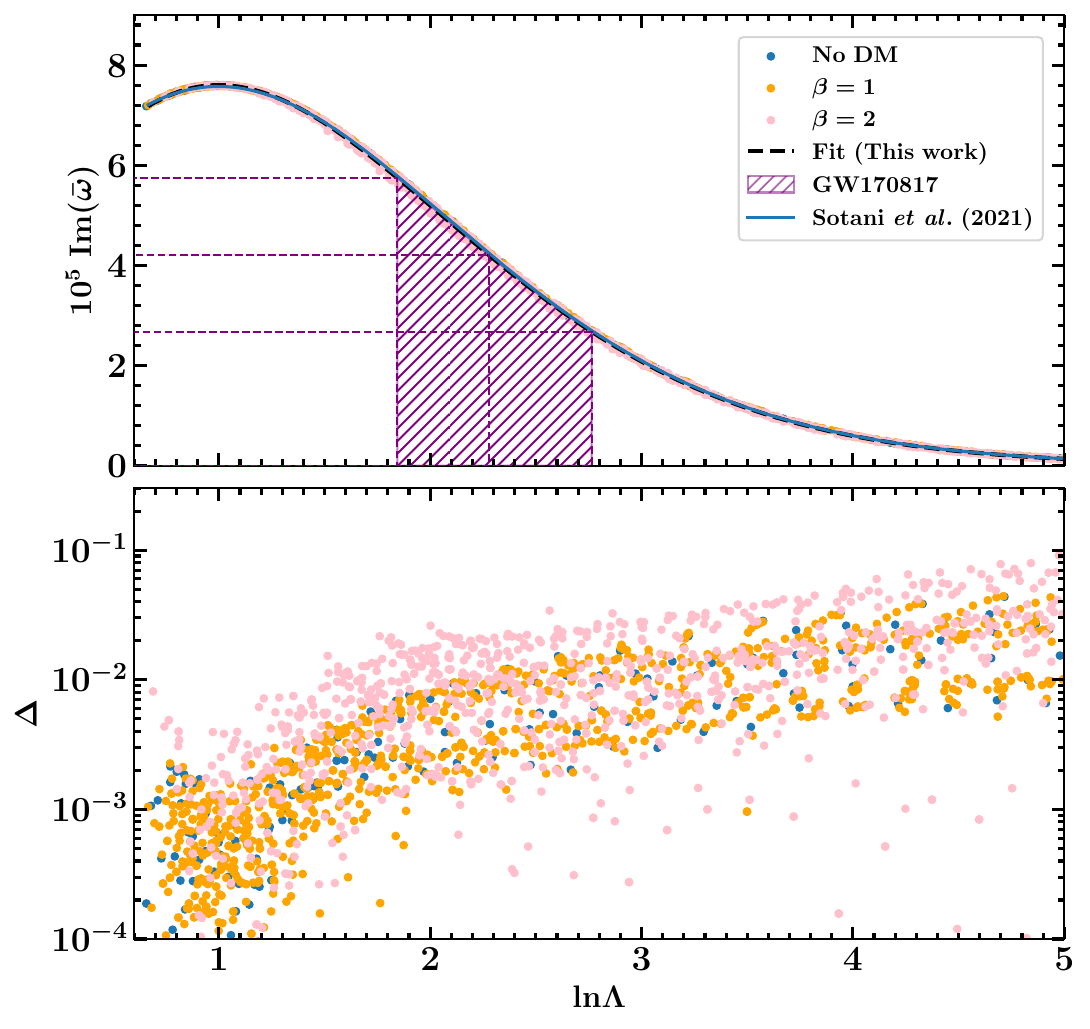}
   \caption{The UR between the scaled QNM frequency and the dimensionless tidal deformability. The black dashed curve shows the least-squares fit from Eq. (\ref{eq:f-Love_fitting}). Fits from Chan {\it et al.} (2014) \cite{Chan_2014}, Sotani {\it et al.} (2021) \cite{Sotani_2021}, Zhao {\it et al.} (2022) \cite{Zhao_2022}, and Bikram {\it et al.} (2022) \cite{Pradhan_2022}  are also included for comparison. Left and right panels display the real and imaginary parts of the QNM frequency, respectively. The observational bounds on tidal deformability from the multimessenger event GW170817 \cite{Abbott_2018} are imposed to constrain the $f$-mode frequency and $\tau$ for the canonical NS model.}
   \label{fig:UR_Lam}
\end{figure*}
In Fig. \ref{fig:UR_Lam}, the universal relation between the scaled QNM frequency and the dimensionless tidal deformability ($\bar{\omega}-\Lambda$) is illustrated. The left panel shows the real part, while the right panel presents the imaginary part of the QNM frequency as functions of $\log(\Lambda)$. The fits incorporate both nucleonic and DM degrees of freedom using Eq. (\ref{eq:f-Love_fitting}). For comparison, previous fitting relations from the literature are also included. The corresponding fitting coefficients, $b_n$ and $b^{\prime}_n$, for the present and earlier works are summarized in Table \ref{tab:f-Love_coef}.  Here, the calculated RMSE and maximum residual for real (imaginary) part are 0.010 (0.028) and 0.066 (0.294) respectively.. The tidal deformability constraint from the multimessenger event GW170817 ($\Lambda_{1.4}=190^{+390}_{-120}$) for the canonical model has been used here. In the current fit, the DM-admixed models are included and we assert that here multimessenger constraint does not limits the DM parameter space ($\alpha M_\chi$, $\beta$). Instead, it used to solely constraints the quadrupole $f$-mode frequency and $\tau$ of canonical NS. The inferred limits from the $\bar{\omega}-\Lambda$ relation obtained in this work are $f_{1.4}=2.110^{+0.445}_{-0.445}$ kHz and $\tau_{1.4}=0.164^{+0.093}_{-0.044}$ s, respectively.

\begin{table*}
    \centering
    \caption{The fitting coefficients are listed for $\bar{\omega}-\Lambda$ relation following the approximate formula Eq. (\ref{eq:f-Love_fitting}). The upper and lower tables show the fitting coefficients for the real and imaginary parts of the QNM frequency, respectively.}
    \setlength{\tabcolsep}{1.7mm}
    \renewcommand{\arraystretch}{1.3}
    \scalebox{0.8}{
        \begin{tabular}{ccccccccc}
            \hline \hline
            Ref. & $a_{0} \left( 10^{-1} \right)$ & $a_{1}\left(10^{-2}\right)$& $a_{2}\left(10^{-2}\right)$ & $a_{3}\left(10^{-3}\right)$ & $a_{4}\left(10^{-4}\right)$& $a_{5}\left(10^{-5}\right)$& $a_{6}\left(10^{-5}\right)$ & $a_{7}\left(10^{-6}\right)$ \\
            \hline \vspace{-3mm} \\
            \parbox[c]{0.15\linewidth}{\centering Chan  {\it et al.} (2014) \cite{Chan_2014}} & $1.820$ & $-1.574$  & $-2.225$ & $6.366$ & $-5.220$ & $--$ & $--$ & $--$ \vspace{2mm} \\
            \parbox[c]{0.15\linewidth}{\centering Sotani {\it et al.} (2021) \cite{Sotani_2021}} & $1.845$ & $-2.074$  & $-1.965$ & $6.256$ & $-7.017$ & $2.780$ & $--$ & $--$ \vspace{1mm} \\
            \parbox[c]{0.15\linewidth}{\centering Zhao {\it et al.} (2022) \cite{Zhao_2022}} & $1.817$ & $-1.532$  & $-2.176$ & $4.971$ & $4.812$ & $-31.040$ & $4.230$ & $-1.971$ \vspace{1mm} \\
            \parbox[c]{0.15\linewidth}{\centering Bikram {\it et al.} (2022) \cite{Pradhan_2022}} & $1.814$ & $-1.341$  & $-2.505$ & $7.736$ & $-8.070$ & $2.038$ & $--$ & $--$ \vspace{1mm} \\
            This work & $1.810$ & $-1.194$  & $-2.713$ & $8.729$ & $-9.300$ & $-1.553$ & $1.016$ & $-0.557$ \vspace{1mm} \\
            \hline \hline
        \end{tabular}} \\ \vspace{0.2cm}
    \scalebox{0.8}{
        \begin{tabular}{ccccccccc}
            \hline \hline
            Ref. & $a^{\prime}_{0}$ & $a^{\prime}_{1}\left(10^{-1}\right)$& $a^{\prime}_{2}\left(10^{-1}\right)$ & $a^{\prime}_{3}\left(10^{-1}\right)$ & $a^{\prime}_{4}\left(10^{-3}\right)$& $a^{\prime}_{5}\left(10^{-5}\right)$& $a^{\prime}_{6}\left(10^{-4}\right)$ & $a^{\prime}_{7}\left(10^{-5}\right)$ \\
            \hline \vspace{-3mm} \\
            \parbox[c]{0.15\linewidth}{\centering Sotani {\it et al.} (2021) \cite{Sotani_2021}} & $-4.334$ & $4.589$  & $-2.795$ & $0.364$ & $-2.518$ & $6.257$ & $--$ & $--$ \vspace{1mm} \\
            This work & $-4.391$ & $6.373$  & $-4.848$ & $1.532$ & $-39.315$ & $662.212$ & $-6.232$ & $2.465$ \vspace{1mm} \\
            \hline \hline
        \end{tabular}}
    \label{tab:f-Love_coef}
\end{table*}

\section{Relevance for current and future GW detectors}
 
To discuss the detectability, we have follow the following formalism \cite{Kokkotas_GW_2001, Flores_GW_2018}, 
\begin{eqnarray}
    \frac{E_{GW}}{M_\odot c^2} = 3.471 \times 10^{36} \bigg(\frac{S}{N}\bigg)^2 \bigg(\frac{1+4Q^2}{Q^2} \bigg) \bigg(\frac{D}{10 \ \rm kpc} \bigg)^2 \bigg(\frac{f}{1 \ \rm kHz} \bigg)^2 \bigg(\frac{S_n}{1 \ \rm kHz^{-1}} \bigg)
\end{eqnarray}
We determine the minimal energy that needs to be released in order to attain a signal-to-noise ratio higher than 5. Here $Q=\pi  f
\tau$ represents the quality factor where $f$ denotes the oscillation frequency of $f$-mode and $\tau$ denotes the damping time. The spectral noise density is indicated by $S_n$. Here we have calculated the GW energy $E_{GW}$ for the canonical mass choices.
\\
We explore two groups of GW detectors. The first illustrates the usual efficiency of Advanced LIGO and Virgo in the kHz frequency range, with a sensitivity of around $S_n^{1/2}=2 \times 10^{-23} \ {\rm Hz^{-1/2}}$ \cite{Abbott_2017}. The second represents the anticipated sensitivity of upcoming third-generation detectors, such the Einstein Telescope, which is anticipated to achieve around $S_n^{1/2}=10^{-24} \ {\rm Hz^{-1/2}}$ in a comparable frequency range \cite{Abbott_2017_ET}. Concerning source distances, we examine two illustrative scenarios: one with a NS positioned in the Virgo cluster at an approximate distance of $D \sim 15 \ {\rm Mpc}$, and another placed inside our Galaxy at around$D \sim 10 \ {\rm kpc}$.

\begin{table*}[htbp]
\centering
\caption{With different DM model, the GW energy is shown for $1.4 \ M_\odot$ of Hornick3 EOS.}
\begin{tabular}{|c|c|c|c|c|c|}
\hline
\multicolumn{6}{|c|}{$1.4\,M_{\odot}$ Configuration} \\
\hline
Distance & Model & $f_{\text{mode}}$ (kHz) & $\tau$ (s) & $E_{\text{GW}}/(M_\odot c^2)$  & $E_{\text{GW}}/(M_\odot c^2)$ \\
&($\alpha M_\chi, \beta$)&&& ($S_n = 2 \times 10^{-23}~\mathrm{Hz}^{-1}$) &  ($S_n = 1 \times 10^{-24}~\mathrm{Hz}^{-1}$) \\
\hline
\multirow{6}{*}{10 kpc}
& (0.01, 1)   & 1.62 & 0.261 & $9.10 \times 10^{-8}$ & $2.27 \times 10^{-10}$ \\
& (0.05, 1)   & 1.66 & 0.255 & $9.56 \times 10^{-8}$ & $2.39 \times 10^{-10}$ \\
& (0.1, 1)    & 1.70 & 0.249 & $1.00 \times 10^{-7}$ & $2.50 \times 10^{-10}$ \\
& (0.01, 2)   & 1.63 & 0.260 & $9.21 \times 10^{-8}$ & $2.30 \times 10^{-10}$ \\
& (0.05, 2)   & 1.68 & 0.252 & $9.79 \times 10^{-8}$ & $2.44 \times 10^{-10}$ \\
& (0.1, 2)    & 1.76 & 0.242 & $1.07 \times 10^{-7}$ & $2.68 \times 10^{-10}$ \\
\hline
\multirow{6}{*}{15 Mpc}
& (0.01, 1)   & 1.62 & 0.261 & $0.20$ & $5.12 \times 10^{-4}$ \\
& (0.05, 1)   & 1.66 & 0.255 & $0.21$ & $5.37 \times 10^{-4}$ \\
& (0.1, 1)    & 1.70 & 0.249 & $0.22$ & $5.64 \times 10^{-4}$ \\
& (0.01, 2)   & 1.63 & 0.260 & $0.20$ & $5.18 \times 10^{-4}$ \\
& (0.05, 2)   & 1.68 & 0.252 & $0.22$ & $5.50 \times 10^{-4}$ \\
& (0.1, 2)    & 1.76 & 0.242 & $0.24$ & $6.04 \times 10^{-4}$ \\
\hline
\end{tabular}
\label{tab:E_GW}
\end{table*}

Table \ref{tab:E_GW} illustrates the GW energy $E_{GW}$ emitted in the $f$-mode by neutron stars with $1.4 \ M_\odot$ configurations across several DM configurations. The varying frequency and damping time at a distance of 10 kpc indicate that the anticipated energy is below the average energy emitted in a supernova ($10^{-5}-10^{-6} \ M_\odot c^2$), rendering detection at this distance plausible. At 15 Mpc, the necessary energy escalates significantly to around $0.20-0.24 \ M_\odot c^2$, far beyond anticipated emission levels, therefore making such remote signals undetected with existing sensitivity capabilities. At a spectral noise density of $S_n = 1 \times 10^{-24}~\mathrm{Hz}^{-1}$, the estimated energy necessary for the detection of GW is around $10^{-10} \ M_\odot c^2$ at a distance of 10 kpc, while at 15 Mp, it varies from $5.12 \times 10^{-4}$ to $6.04 \times 10^{-4} \ M_\odot c^2$. We also examine how the microphysical alterations in the EOS, particularly via the DM, influence the GW energy released by NSs. The incorporation of DM softens the EOS, resulting in significantly more compact stars characterized by increased mode frequencies and reduced damping times. These modifications augment the GW luminosity and shift the $f$-mode signal into greater detectable ranges. Consequently, the DM components reveal clearly identifiable patterns in the GW energy production, underscoring their significance for forthcoming asteroseismic studies using advanced detectors.

\section{Summary and Conclusion}
\label{summary}
In this work, a comprehensive investigation of nonradial oscillations in DM admixed NS is performed. The relativistic mean field formalism has been utilized to model the hadronic matter EOS. A physically plausible Higgs portal DM model is employed to describe the system, wherein the DM distribution inside the NS is non-uniform and governed by gravitational potential. To regulate the influence of DM on the stellar properties, two parameters are introduced: $\alpha$ and $\beta$. This study combines the latest advances in NS microphysics, general relativistic perturbation theory, and multimessenger constraints to elucidate how DM alters both equilibrium structure and dynamical oscillation properties. By scanning a two-parameter family of DM models ($\alpha M_\chi$, $\beta$), the effects of DM on mass-radius relations, tidal deformability, $f$-mode frequency and $\tau$ have been explored. Furthermore, correlations among these quantities are explored for the DM-admixed configurations, and the URs are derived in the presence of DM.

The presence of DM softens the overall EOS, leading to a reduction in both the maximum stable mass and radius of the NS. As the parameters $\alpha M_\chi$ and $\beta$ increase, this effect becomes more pronounced, causing the mass-radius relation to deviate from observational bounds when a significant DM component is present. The maximum DM fraction, $f_{\chi , max}$ within the NS has also been computed, revealing its dependence on both $\alpha M_\chi$ and $\beta$. For very small $\alpha M_\chi$, the $f_{\chi , max}$ remains negligible regardless of $\beta$. However, as $\alpha M_\chi$ increases, the $f_{\chi , max}$ grows for moderate $\beta$ values but decreases at high $\beta=4$, suggesting that an excessively steep profile suppresses the DM contribution. These findings highlight the importance of exploring the parameter space of ($\alpha M_\chi$, $\beta$) to limit the $f_{\chi , max}$ within NSs that remain consistent with current astrophysical mass-radius constraints.

Parallel to the analysis of static NS structure, this work also investigates the non-radial oscillations of DM-admixed NS using general relativistic perturbation theory to compute the complex QNM frequencies. The $f$-mode and its corresponding $\tau$ ($\tau$) are evaluated to probe the dynamical response of the star. The analysis begins by examining how the inclusion of DM, characterized by the parameters $\alpha M_\chi$ and $\beta$, influences the $f$-mode frequency. It is observed that the presence of the DM significantly shifts the $f$-mode frequency to higher value for a given mass as compared to without DM. This behavior holds true even when the constant tidal deformability is considered. The effect become more noticeable for the NS that are more massive or have lower tidal deformability. In contrast, the $\tau$ exhibits the opposite behavior, decreasing in the presence of DM. Just like the $f$-mode frequency, the change in the $\tau$ is more significant for NS with high mass and lower tidal deformability. To further understand this phenomenon, both the $f$-mode frequency and the corresponding $\tau$ are analyzed as functions of $\alpha M_\chi$ and $f_{\chi , max}$ across different stellar models. The results show that as $\alpha M\chi$ and $f_{\chi , max}$ increase, the $f$-mode frequency rises while the $\tau$ decreases. A similar behavior is observed when examining the canonical $f$-mode frequency and $\tau$ as functions of $\alpha M_\chi$ and central energy density, the frequency increases with higher central density, while the $\tau$ decreases. This implies that DM-rich NSs tend to oscillate at higher frequencies and experience enhanced damping, primarily due to the stronger coupling between matter and spacetime perturbations induced by DM. 

Regardless of examining how the presence of DM influences NS observables and QNM characteristics, this work also explores correlation among the DM model parameters, NS observables, and $f$-mode properties. The results demonstrates that the incorporation of DM degrees of freedom, while influencing the internal composition of NSs, preserves the fundamental correlations among global observables and $f$-mode properties. The tight correlation between tidal deformability, radius, and oscillation parameters remain consistent with established scaling relations, indicating the robustness of these relations even in the presence of DM. The perfect correlation between the DM effective controlling parameter and the corresponding DM mass fractions further reflects a coherent DM influence across stellar configurations. Altogether, these results suggest that $f$-mode asteroseismology, when combined with future multimessenger observations, could offer valuable constraints on the properties of DM and its interaction with dense baryonic matter.

Finally, the URs within the framework of DM-admixed NS asteroseismology are examined. The $f$-mode frequency is fitted as a function of the average stellar density, while the scaled $\tau$ is correlated with the stellar compactness. The obtained fits are compared with those from previous studies, and the corresponding fitting coefficients are reported. Moreover, the universality between the stellar compactness and the mass-scaled $f$-mode frequency, as well as between compactness and the scaled $\tau$, is analyzed. The results reveal that the inclusion of DM does not break these universal relations, indicating their robustness even in the presence of DM-induced modifications to the stellar structure. Furthermore, the universality between the dimensionless tidal deformability and both the scaled $f$-mode frequency and $\tau$ is explored. Using the tidal deformability constraint from the multimessenger GW event GW170817, the present analysis constrains the $f$-mode frequency and $\tau$ for a canonical NS to $f_{1.4}=2.110^{+0.445}_{-0.445}$ kHz and $\tau_{1.4}=0.164^{+0.093}_{-0.044}$ s, respectively. These results will enable future GW observations to infer QNM frequencies through the established URs. Since DM influences both the static and dynamical properties of NSs, extracting observational signatures from such complex systems remains a significant challenge for future studies.

\section*{ACKNOWLEDGMENTS}
I would like to thank my supervisor Dr. Bharat Kumar for his valuable discussions and insightful suggestions throughout this work. I also want to thank Mr. Probit J. Kalita for the discussion during the project.

\appendix

\section{Hadronic equation of state}
\label{Appendix-A}
In the present study, the hadronic EOSs are constructed within the framework of the RMF formalism. In this approach, nucleons interact through the exchange of mesons, while the mesons themselves are allowed to exhibit self-interactions and cross-couplings. The model includes three types of mesons: the scalar $\sigma$, the vector $\omega$, and the isovector $\rho$ mesons. The total Lagrangian density for NS matter is composed of contributions from both the nucleonic and leptonic ($e^-$ and $\mu^-$) sectors, and can be expressed as \cite{FURNSTAHL_1996, Singh_2014, Kumar_2018, pinku_jcap_2023, Probit_2024}:
\begin{equation}
    \begin{aligned}
        \mathcal{L}_{\text{NS}} &= \sum_{\alpha=n,p} \bar{\psi}_{\alpha} \left[ \gamma_{\mu} \left( i\partial^{\mu} - g_{\omega} \omega^{\mu} - \frac{1}{2} g_{\rho} {\tau} \cdot {\rho}^{\mu} \right) \right. \\
        & \left. - \left( M_N -\vphantom{\frac{a}{b}} g_{\sigma} \sigma \right) \right] \psi_{\alpha} + \frac{1}{2}\partial_{\mu}\sigma\partial^{\mu}\sigma - \frac{1}{2}m_{\sigma}^2\sigma^2 \\
        & +\frac{\zeta_0}{4!}g_\omega^2(\omega^{\mu}\omega_{\mu})^2-\frac{\kappa_3}{3!}\frac{g_{\sigma}m_{\sigma}^2\sigma^3}{M_N}-\frac{\kappa_4}{4!}\frac{g_{\sigma}^2m_{\sigma}^2\sigma^4}{M_N^2} \\
        & + \frac{1}{2}m_\omega^2\omega_{\mu}\omega^{\mu} - \frac{1}{4}W^{\mu \nu}W_{\mu\nu} + \frac{1}{2}m_\rho^2{\rho^\mu}\cdot{\rho_\mu} \\
        & - \frac{1}{4}{R^{\mu\nu}}\cdot{R_{\mu\nu}} - \Lambda_\omega g_\omega^2 g_\rho^2 (\omega^\mu \omega_\mu)({\rho^\mu}\cdot{\rho_\mu)} \\
        & + \sum_{l=e, \mu} \bar{\psi}_k (i\gamma^\mu \partial\mu - m_k) \psi_k
    \end{aligned}
    \label{eq:lagrangian}
\end{equation}
Here, $\psi$ denotes the nucleonic Dirac spinor, while $\sigma$, $\omega^\mu$, and $\rho^\mu$ represent the scalar ($\sigma$), vector ($\omega$), and vector–isovector ($\rho$) meson fields, respectively. The quantities $M_N$, $m_\sigma$, $m_\omega$, and $m_\rho$ correspond to the masses of the nucleons and mesons. The parameters $g_\sigma$, $g_\omega$, and $g_\rho$ are the meson–nucleon coupling constants, whereas $\kappa_3$ and $\kappa_4$ denote the third- and fourth-order self-coupling coefficients of the scalar meson field. The constants $\zeta_0$ and $\Lambda_\omega$ account for the vector meson self-interaction and the vector–isovector meson coupling, respectively. The antisymmetric field tensors are defined as $W_{\mu\nu} = \partial_\mu \omega_\nu - \partial_\nu \omega_\mu$ and $R_{\mu\nu} = \partial_\mu \rho_\nu - \partial_\nu \rho_\mu$, and $\tau$ represents the isospin operator. In addition, $\psi_k$ and $m_k$ denote the leptonic Dirac spinor and mass, respectively.

The hadronic EOSs used in this model utilized the above lagrangian density for their calculation. The parameter for the current EOSs shown in the Tab. \ref{tab:parameter}. 
\begin{table*}
\centering
\caption{The masses of the $\sigma$, $\omega$, and $\rho$ mesons are in MeV. The nucleon mass ($M_N$) is taken as 939 MeV.}
\renewcommand{\tabcolsep}{0.35cm}
\renewcommand{\arraystretch}{1.3}
\scalebox{0.75}{
\begin{tabular}{lcccccccccc}
\hline \hline
Model & $m_\sigma$ & $m_\omega$ & $m_\rho$ & $g_\sigma$   & $g_\omega$   &$g_\rho$   & $\kappa_3$   &$\kappa_4$ & $\Lambda_{\omega}$ & $\zeta_0$\\
\hline
 QMC-RMF4 \cite{Alford_2022} & $491.5$ & $782.5$ & $763$ & $8.21$ & $9.94$ & $12.18$ & $2.01736$ & $-3.09985$ & $0.1055$ & $0$\\
 NITR-I \cite{pinku_ijmpe_2024}   & $470$ & $782.5$ & $763$ & $8.729$ & $11.172$ & $9.461$ & $2.729$ & $-10.207$ & $0.029$ & $0.159$\\
 Hornick1 \cite{Hornick_2018} & $550$ & $783$ & $770$ & $11.552$ & $13.566$  & $11.871$ & $1.5471$ & $-6.6419$ & $0.0295$ & $0$\\
 Hornick2 \cite{Hornick_2018} & $550$ & $783$ & $770$ & $10.993$ & $12.708$  & $10.651$ & $1.6729$ & $-6.6740$ & $0.0272$ & $0$\\
 Hornick3 \cite{Hornick_2018} & $550$ & $783$ & $770$ & $10.429$ & $11.774$  & $10.186$ & $1.9556$ & $-7.0031$ & $0.0278$ & $0$\\
 Hornick4 \cite{Hornick_2018} & $550$ & $783$ & $770$ & $9.846$  & $10.746$  & $9.982$  & $2.4385$ & $-7.3696$ & $0.0314$ & $0$\\
\hline \hline
\end{tabular}}
\label{tab:parameter}
\end{table*}

\bibliography{dm}

\end{document}